\documentclass{aims} 
\usepackage{amsmath}
\usepackage{amssymb}

\usepackage{paralist}
\usepackage[misc]{ifsym}
\usepackage{booktabs}
\usepackage{multirow}
\usepackage{subcaption}  
\usepackage{epsfig} 
\usepackage{epstopdf} 

\newcommand{\bfA}{{\bf A}}

\newcommand{\bfD}{{\bf D}}

\newcommand{\bfI}{{\bf I}}

\newcommand{\bfU}{{\bf U}}

\newcommand{\bfb}{{\bf b}}

\newcommand{\bfs}{{\bf s}}

\newcommand{\bfx}{{\bf x}}

\newcommand{\bfu}{{\bf u}}

\newcommand{\bfv}{{\bf v}}

\newcommand{\bfz}{{\bf z}}
\newcommand\sitd{{DAW-FM}}
\newcommand\sitdn{{DAWN-FM}}

\newcommand{\bfepsilon}{{\boldsymbol \varepsilon}}
\newcommand{\bftheta}{{\boldsymbol \theta}}

\newcommand{\bfeta}{{\boldsymbol \eta}}

\newcommand{\bflambda}{{\boldsymbol \lambda}}

\newtheorem{example}{Example}[section]

\usepackage[colorlinks=true]{hyperref}
\hypersetup{urlcolor=blue, citecolor=red}
\allowdisplaybreaks
\usepackage{cleveref}

\def\currentvolume{X}
 \def\currentissue{X}
  \def\currentyear{200X}
   \def\currentmonth{XX}
    \def\ppages{X-XX}
     \def\DOI{10.3934/xx.xxxxxxx}

\theoremstyle{definition}

\title[Data-Aware and Noise-Informed Flow Matching]
{DAWN-FM: Data-Aware and Noise-Informed Flow Matching for Solving Inverse Problems} 

\author[Shadab Ahamed and Eldad Haber]{}

\subjclass{Primary: 65J20, 65J22, 68T07; Secondary: 65C30.}
\keywords{Inverse problems, Flow matching, Stochastic interpolation, Image deblurring, Tomography, Data embedding, Noise embedding}

\thanks{The first author is supported by Canadian Institutes of Health Research (CIHR)
Grant PIBH-GR018169.}

\thanks{$^*$Corresponding author: Shadab Ahamed}

\begin{document}
\maketitle

\centerline{\scshape
Shadab Ahamed$^{{\href{mailto:shadab.ahamed@hotmail.com}{\textrm{\Letter}}}*1}$
and Eldad Haber$^{{\href{mailto:ehaber@eoas.ubc.ca}{\textrm{\Letter}}}2}$}

\medskip

{\footnotesize
 \centerline{$^1$Department of Physics \& Astronomy, University of British Columbia, Vancouver, BC Canada}
} 

\medskip

{\footnotesize
 \centerline{$^2$Department of Earth, Ocean and Atmospheric Sciences, University of British Columbia, Vancouver, BC Canada}
}

\bigskip

 \centerline{(Communicated by Carola-Bibiane Sch\"onlieb)}


\begin{abstract}
Inverse problems, which involve estimating parameters from incomplete or noisy observations, arise in various fields such as medical imaging, geophysics, and signal processing. These problems are often ill-posed, requiring regularization techniques to stabilize the solution. In this work, we employ \textit{Flow Matching (FM)}, a generative framework that integrates a deterministic processes to map a simple reference distribution, such as a Gaussian, to the target distribution. Our method \textit{\textbf{DAWN-FM}}: \textit{\textbf{D}ata-\textbf{AW}are and \textbf{N}oise-Informed \textbf{F}low \textbf{M}atching} incorporates \textit{data and noise embedding}, allowing the model to access representations about the measured data explicitly and also account for noise in the observations, making it particularly robust in scenarios where data is noisy or incomplete. By learning a time-dependent velocity field, FM not only provides accurate solutions but also enables uncertainty quantification by generating multiple plausible outcomes. Unlike pretrained diffusion models, which may struggle in highly ill-posed settings, our approach is trained specifically for each inverse problem and adapts to varying noise levels. We validate the effectiveness and robustness of our method through extensive numerical experiments on tasks such as image deblurring and tomography. The code is available at: \url{https://github.com/ahxmeds/DAWN-FM.git}.
\end{abstract}


\section{Introduction}

Inverse problems are a class of problems in which the goal is to determine parameters (or parameter function) of a system from observed  data. These problems arise in various fields, including medical imaging, geophysics, remote sensing, and signal processing. Inverse problems are often ill-posed, meaning that a unique solution does not exist, or the solution may be highly sensitive to small perturbations in the data. 
Solving inverse problems typically requires regularization techniques, which introduce additional constraints or prior information to stabilize the solution and mitigate the effects of ill-posedness.
Such regularization can be obtained using various mathematical and computational methods, including optimization techniques (variational methods), statistical inference (Bayesian or frequentist) and machine learning \cite{Scherzer2009, adler2017solving, calvetti2024distributed, LOPEZTAPIA2021103285}. In this paper, we approach inverse problems using the latter and investigate a new set of methods that are machine learning based for the solution of inverse problems. In particular, we show how flow matching (FM) which was recently proposed by Lipman et al.~\cite{lipman2022flow} in the context of generative models can be effectively used to estimate the solution of inverse problems and to further investigate its non-uniqueness.

Stochastic Interpolation (SI) is a relatively new generative process that provides a unifying framework, elegantly integrating both deterministic flows and stochastic diffusion models \cite{albergo2022building,albergo2023stochastic}. The core concept of SI is to learn a stochastic process that effectively transports a simple reference distribution, such as Gaussian, to the desired target data distribution. This transportation process can manifest as either deterministic or stochastic. In the former case, it is described by an ordinary differential equation (ODE), while in the latter, it is governed by a stochastic differential equation (SDE). For this work, we rely only on the formal deterministic ODE formalism of SI, which is known as Flow Matching (FM).

The FM framework defines a continuous-time reversibility between the reference and target distributions, parameterized by time $t \in [0,1]$. At the initial time $t=0$, the distribution aligns with the reference distribution. As time progresses to $t=1$, the distribution evolves to match the target data distribution. This evolution is achieved by learning the time-dependent velocity (or drift) field that characterize this interpolation process. By understanding and modeling this time-dependent transformation, one can generate samples from the target distribution through numerical integration of the learned ODE.

FM is a highly flexible methodology for designing new types of generative models. In this work, we use this flexibility and show how to adopt FM to solve highly ill-posed inverse problems. We find FM particularly useful since it allows ease of sampling from the target distribution. This implies that we are able to generate a range of solutions to the inverse problems and thus investigate the uncertainty that is associated with the estimated solution. Such a process is highly important in physical applications when decisions are made based on the solution.

\textbf{Related work:} The methods proposed here are closely related to three different approaches for the solution of inverse problems. First, several studies have incorporated diffusion models as regularizers in inverse problems \cite{yang2022diffusion,darassurvey}. Chung et al.~\cite{chung2022diffusion} proposed diffusion posterior sampling strategy to efficiently solve noisy non-linear inverse problems via approximation of the posterior sampling. Chung et al.~\cite{chung2022come} introduced Come-Closer-Diffuse-Faster technique to show how the feed-forward neural networks for inverse problems can be synergistically combined with the diffusion models for a variety of problems like image inpainting and super-resolution. Song et al.~\cite{song2022solving} proposed a fully unsupervised method for inverse problems based on score-matching diffusion models, which reconstructs an image consistent with both the prior and the observed measurements. Li et al.~\cite{li2022srdiff} proposed SRDiff, a diffusion-based model that generates high-resolution images from low-resolution inputs by formulating single image super-resolution as a conditional diffusion process. Whang et al.~\cite{whang2022deblurring} introduced a technique for deblurring using conditional diffusion models which trains a stochastic sampler that refined the output of a deterministic predictor, producing a diverse set of plausible reconstructions for a given
input, thereby improving image perceptual quality. The key idea often used in these methods was to leverage a pretrained diffusion model that captures the data distribution as a prior, and then condition this model on the given measurements to infer the underlying clean signal or image. For inverse problems, diffusion methods condition the diffusion model on the given measurements (e.g. noisy, incomplete, or compressed data) by incorporating them into the denoising process. Nonetheless, it has been shown in \cite{eliasof2024completedeeplearningmethod} that pretrained diffusion models that are used for ill-posed inverse problems as regularizers tend to under-perform as compared to the models that are trained specifically on a particular inverse problem. In particular, such models tend to break when the noise level is not very low.

A second branch of techniques that are related to the work proposed here uses encoder–decoder style neural architectures for solving inverse problems. For instance, Chung et al.~\cite{chung2024pairedautoencodersinverseproblems} proposed paired-autoencoders for solving inverse problems that learns joint latent representations of data and model spaces, enabling likelihood-free estimation for PDE-based inverse problems while also providing refinement strategies. Radev et al.~\cite{radev2020bayesflow} proposed an invertible neural network combined with a learned summary (encoder) network to map observations to latent parameters, enabling amortized posterior inference without explicit likelihoods. Wang et al.~\cite{wang2021theory} proposed theory-guided autoencoder which discretized PDE residuals into an autoencoder loss, enabling surrogate modeling and inverse parameter estimation for subsurface flow with limited labeled data. Padmanabha et al.~\cite{padmanabha2021solving} constructed a conditional invertible neural network that maps measurements to latent variables (and vice versa) under a learned bijective transformation, allowing likelihood-free inversion by running the invertible mapping backward to recover parameter estimates conditioned on observations. Together, these works illustrate how encoder–decoder architectures can be harnessed as powerful likelihood-free estimators, enabling direct data-to-solution mappings that bypass explicit likelihood evaluation while tackling complex inverse problems. This property is especially advantageous when the forward operator is either computationally intractable or analytically unavailable, which is often the case in high-dimensional imaging applications.

A third branch of techniques that relates to our approach integrates the forward problem directly into the neural network architecture. In this line of work, the forward operator is embedded into the training loop so that the data misfit (and its gradient) can be computed within the network and used as an explicit guide for learning. For instance, Eliasof et al.~\cite{eliasof2024completedeeplearningmethod} introduced InverseUNetODE, which incorporates the adjoint of the forward model at each layer of a UNet to iteratively refine reconstructions, demonstrating improved stability in highly ill-posed settings. Mukherjee et al.~\cite{mukherjee2021learning} proposed learning convex regularizers that satisfy variational source conditions, combining deep learning with principles of classical variational analysis to enforce forward-consistency during inversion. Similarly, Eliasof et al.~\cite{eliasof2023drip} introduced a method called DRiP, a variational framework that learns neural regularization functionals guaranteeing data consistency by combining least-action principles with deep neural architectures, thereby overcoming the instability of proximal-based methods. Jin et al.~\cite{jin2017deep} introduced FBPConvNet, which combines filtered back projection with a U-Net-style CNN to remove artifacts from ill-posed inverse reconstructions, effectively embedding the forward operator into the learning process to ensure physics-consistent solutions.
Collectively, these methods highlight the value of embedding the forward model within the network, as they not only reduce the reliance on external priors but also ensure that learned reconstructions remain anchored to the measured data. Our approach utilizes components from this methodology by incorporating forward-model–based misfit terms into the training of flow-matching networks, thereby grounding the generative process in the measured observations.

Additionally, a complementary line of research has recently focused on guided flow matching techniques, which extend the basic FM formalism by introducing additional control signals or constraints during training and sampling. Zheng et al.~\cite{zheng2023guided} proposed Guided Flows, incorporating classifier-free guidance into velocity fields to achieve conditional generative modeling across images, speech, and decision-making tasks. Chemseddine et al.~\cite{chemseddine2025conditional} introduced conditional Wasserstein distance that aligns posterior distributions by forbidding mass transport in the observation space, thereby enabling Bayesian optimal transport-based flows. Ben-Hamu et al.~\cite{ben2024d} introduced D-Flow, a framework that differentiates through the diffusion/FM flow to enable controlled generation by backpropagating task-specific objectives D-Flow, a framework that controls generation in diffusion and flow-matching models by differentiating through the flow, optimizing for source point. Furthermore, Pokle et al.~\cite{pokle2023training} demonstrated training-free linear image inversion using pretrained flow models, where gradient correction techniques and conditional optimal transport probability paths are used to solve inverse problems. Together, these works illustrate the versatility of guided flow matching, highlighting how auxiliary signals - ranging from classifier-based gradients to transport constraints - can steer the generative dynamics toward desired data distributions or application-specific goals. Similarly, our work also introduces guidance via incorporating measured data as well as noise variance directly in the network architecture and optimizing it using measured data misfit to direct the flow toward solving inverse problems in imaging.

\textbf{Main contribution:} The core contributions of this paper lie in the design and application of the \textbf{D}ata-\textbf{AW}are and \textbf{N}oise-\textbf{I}nformed \textbf{F}low \textbf{M}atching (\sitdn) framework, specifically crafted for solving highly ill-posed inverse problems like image deblurring and tomography. (i) Our method is designed to be problem-specific, which adjusts itself to the unique structure of the inverse problem it is tasked with, learning the posterior distribution directly, ensuring that the learned velocity fields and mappings are directly applicable to the target task. (ii) We train an interpolant by embedding measured data and noise information directly into the interpolation process. This incorporation allows our model to adapt to a wide range of noise conditions, something that pretrained models struggle with. By making the training explicitly aware of the noise level and data characteristics, our model can better navigate noisy or incomplete measured data, producing superior reconstructions. (iii) By learning the posterior distribution directly and leveraging the stochastic nature of the interpolation process, \sitdn~generates multiple plausible solutions for a given inverse problem, allowing us to explore the solution space more thoroughly and in particular, to estimate the posterior mean and its standard deviation, estimating the uncertainty in the recovered solution. 

\textbf{Organization of the paper:} The remainder of this paper is organized as follows. In \Cref{sec:fm_inv_problems}, we review the fundamentals of Flow Matching and describe how it can be adapted to solve inverse problems, including the design of our proposed data-aware and noise-informed velocity estimator. \Cref{sec:inverse_problems_details} provides mathematical details for the two inverse problems we solve in this work -- image deblurring and tomography. \Cref{sec:numerical_experiments} presents numerical experiments on image deblurring and tomography, evaluating the performance of the proposed methods against existing approaches. \Cref{sec:conclusion} concludes the paper with a summary of findings and a discussion on future research directions. Additional technical details, including evaluation metrics, experimental settings and additional visualizations are provided in the Supplementary Material (\Cref{app:evaluation_metrics,app:inverse_crime,app:experiment_setting,app:additional_examples}).

\section{Flow Matching and Inverse Problems}
\label{sec:fm_inv_problems}
In this section we review flow matching as well as derive the main ideas behind using it for the solution of inverse problems.

\subsection{Flow Matching (FM): A partial review}

Flow Matching (FM) is a framework that transforms points between two distributions. Given two densities $\pi_0(\bfx)$ and $\pi_1(\bfx)$, the goal is to find a mapping that takes a point $\bfx_0 \sim \pi_0(\bfx)$ and transports it to a point $\bfx_1 \sim \pi_1(\bfx)$.
For simplicity and for the purpose  of this work, we choose $\pi_0$ to be a Gaussian distribution with $0$ mean and $\bfI$ covariance.

Following the linear interpolation described in \cite{liu2022flow}, we define trajectories  parametrized by the time variable $t$ as,
\begin{equation}
    \label{eq:traj}
    \bfx_t = (1-t) \bfx_0 + t \bfx_1, 
\end{equation}
where $\bfx_t \sim \pi_t(\bfx)$ is the linear interpolation of $\bfx_0$ and $\bfx_1$ at any given time $t$. These trajectories connect points from $\bfx_0$ at $t=0$ to $\bfx_1$ at $t=1$, i.e., the parameter $t \in [0,1]$ interpolates between the reference distribution and target distribution and appears explicitly in the velocity estimator (see \Cref{eq:velint}) to ensure that the learned flow evolves smoothly and consistently along this continuum. More complex trajectories have been proposed in \cite{albergo2023stochastic}, however, in our context, we found that simple linear trajectories suffice and have advantages for being very smooth in time. The velocity along the trajectory is the time derivative of the interpolant $\bfx_t$, that is, 
\begin{equation}
    \label{eq:vel}
    {\frac {d\bfx_t}{dt}} = \bfv = \bfx_1 - \bfx_0. 
\end{equation}

Using the velocity vector defined in \Cref{eq:vel}, we aim drive the flow to follow the direction $(\bfx_1 - \bfx_0)$ of the linear path pointing from $\bfx_0 \sim \pi_0$ to $\bfx_1 \sim \pi_1$. To this end, we parameterize the velocity by a learnable function $\bfs_{\bftheta}(\bfx_t, t)$ and solve the optimization problem for $\bftheta$ as,
\begin{equation}
    \label{eq:velest}
    \hat \bftheta = {\rm arg}\min_{\bftheta} {\mathbb E}_{\bfx_0, \bfx_1, t} \Big[\| \bfs_{\theta}(\bfx_t, t) - \bfv \|^2 \Big] = {\rm arg}\min_{\bftheta} {\mathbb E}_{\bfx_0, \bfx_1, t} \Big[\| \bfs_{\theta}(\bfx_t, t) + \bfx_0 - \bfx_1 \|^2 \Big],
\end{equation}
where $t \sim \mathcal{U}[0,1]$ is a uniform random variable \cite{lipman2022flow,liu2022flow}. After training, the velocity function $\bfs_{\bftheta}(\bfx_t, t)$ can be estimated for every point in time. In this work, given $\bfx_0$, we recover $\bfx_1$ by numerically integrating the ODE
\begin{equation}
    \label{eq:velint}
    {\frac {d\bfx_t}{dt}} = \bfs_{\bftheta}(\bfx_t, t), \quad \bfx(0) = \bfx_0, \quad t\in [0,1],
\end{equation}
by Fourth-Order Runge Kutta method with fixed step size. An alternative version where the samples are obtained by using a stochastic differential equation can also be used.

While it is possible to estimate the velocity $\bfv$ from $(\bfx_t, t)$ and then compare it to the true velocity, it is sometimes easier to work with the velocity as a denoiser network, that is, estimate $\bfx_1$ and use the loss of comparing $\bfx_1$ to its denoised quantity. To this end, note that
\begin{equation}
    \label{eq:vfromxt}
    \hat{\bfx}_1 = \bfx_t + (1-t)\bfv \approx \bfx_t + (1-t)\bfs_{\bftheta}(\bfx_t,t).
\end{equation}
\Cref{eq:vfromxt} is useful when a recovered $\bfx_1$ is desirable. In the context of using FM for inverse problems, obtaining an approximation to $\bfx_1$ can be desirable, as we see next.

\subsection{Applying FM for solving inverse problems}

Consider the case where we have observations on $\bfx_1$ of the form
\begin{equation}
    \label{eq:fp}
    \bfA \bfx_1 + \bfepsilon = \bfb.
\end{equation}
Here, $\bfA$ is a linear forward mapping (although the method can work for nonlinear mappings as well) and $\bfepsilon \sim \mathcal{N}(0, \sigma^2 \bfI)$ is a random vector. We assume that $\bfA$ is rank-deficient or numerically rank-deficient \cite{hansen}, so the effective dimension of $\bfb$ is smaller than $\bfx_1$ and one cannot obtain a reasonable estimate for $\bfx_1$ given the noisy data $\bfb$ without the incorporation of a-priori information.

Using Bayes' theorem, we have
\begin{equation}
    \label{eq:bayes}
    \pi(\bfx_1|\bfb) \propto \pi(\bfb|\bfx_1) \pi(\bfx_1).
\end{equation}
Bayes' theorem suggests that it is possible to factor the posterior distribution $\pi(\bfx_1|\bfb)$ using
the known distribution of $\pi(\bfb|\bfx_1)$ and the prior distribution $\pi(\bfx_1)$. This observation motivated a number of studies that used the estimated pretrained distribution $\pi(\bfx_1)$ in the process of solving an inverse problem \cite{yang2022diffusion, chung2022diffusion, chung2022come, chung2022improving, song2022solving}. Nonetheless, it has been shown in \cite{eliasof2024completedeeplearningmethod} that such estimators tend to produce unsatisfactory results when solving highly ill-posed problems or when the data is very noisy. The reason for this behaviour stems from the fact that pretrained estimators push the solution towards the center of the prior distribution, irrespective of what the data represents. To see this, we use a careful dissection of the solution. Consider the singular value decomposition  
$$\bfA = \sum_i\bflambda_i \bfu_i\bfv_i^{\top},$$
where $\bfu_i$ and $\bfv_i$ are the left and right singular vectors and $\bflambda_i$ are the singular values.
Given the orthogonality of $\bfu_i$, we can decouple the data equations into $$ \bflambda_i (\bfv_i^{\top} \bfx_1) = (\bfu_i^{\top} \bfb), \quad i=1,\ldots,n.$$
If $\bflambda_i$ is large, the projection of $\bfx_1$ onto the eigenvector $\bfv_i$ is very informative and very minimal regularization is required. However, if $\bflambda_i \approx 0$, the contribution of $\bfv_i$ to the solution is difficult, if not impossible, to obtain and this is where the regularization is highly needed. When using pretrained models, the prior $\pi(\bfx_1)$ is estimated numerically and it is unaware of the inverse problem at hand. Errors in the estimated $\pi(\bfx_1)$ in the parts that correspond to the large singular vectors may not be destructive. However, if $\pi(\bfx_1)$ has errors that correspond to the very small singular values, that is, to the effective null space of the data, this may lead to the artifacts that have been observed early in \cite{somersallo, Tenorio2011}.
This suggests that, although it is appealing to use a generic pretrained priors in the process, a better approach is to not use the Bayesian factorization to prior and likelihood but rather train a flow matching interpolant that maps the distribution $\pi(\bfx_0)$ to the posterior $\pi(\bfx_1|\bfb)$ directly. Indeed, as we show in \Cref{subsec:a_data_aware_and_noise_informed_velocity_estimator}, the flexibility of the FM framework allows us to learn a velocity function that achieves just that.

\textbf{Inverse crime considerations.} In the original formulation of our inverse problem setup, the measured data $\bfb$ is generated via $\bfb = \bfA \bfx_1 + \bfepsilon$, where $\bfA$ is the same forward operator subsequently used during inversion. This formulation, while standard in many machine learning-based inverse problem studies, can inadvertently lead to an inverse crime \cite{wirgin2004inverse} - a situation where the forward model used for synthetic data generation is identical to that used for inversion, potentially yielding overly optimistic reconstruction results. To rigorously assess the robustness of our proposed DAWN-FM framework under more realistic conditions, we performed additional experiments (reported in \Cref{app:inverse_crime} under Supplementary Material) in which the forward model used for data generation was intentionally modified. Specifically, we employed a composite operator $\Tilde{\bfA} = \bfD \bfA \bfU$ where $\bfD$ denotes a downsampling operator, $\bfU$ an upsampling operator, with $\bfA$ being the original forward model. This change ensures that the forward mapping used to synthesize the training and test data differs from that used during inversion, thereby eliminating the risk of inverse crime. Our results confirm that the effectiveness of our method does not depend on any implicit inverse crime, and the method generalizes well even when the forward model used for data generation is perturbed. For detailed discussion and results considering the issue of inverse crime, refer to \Cref{tab:resolveIC} under \Cref{app:inverse_crime} in the Supplementary Material. 


\subsection{A data-aware and noise-informed velocity estimator}
\label{subsec:a_data_aware_and_noise_informed_velocity_estimator}
\begin{figure}[h]
\centering
\includegraphics[width=\textwidth]{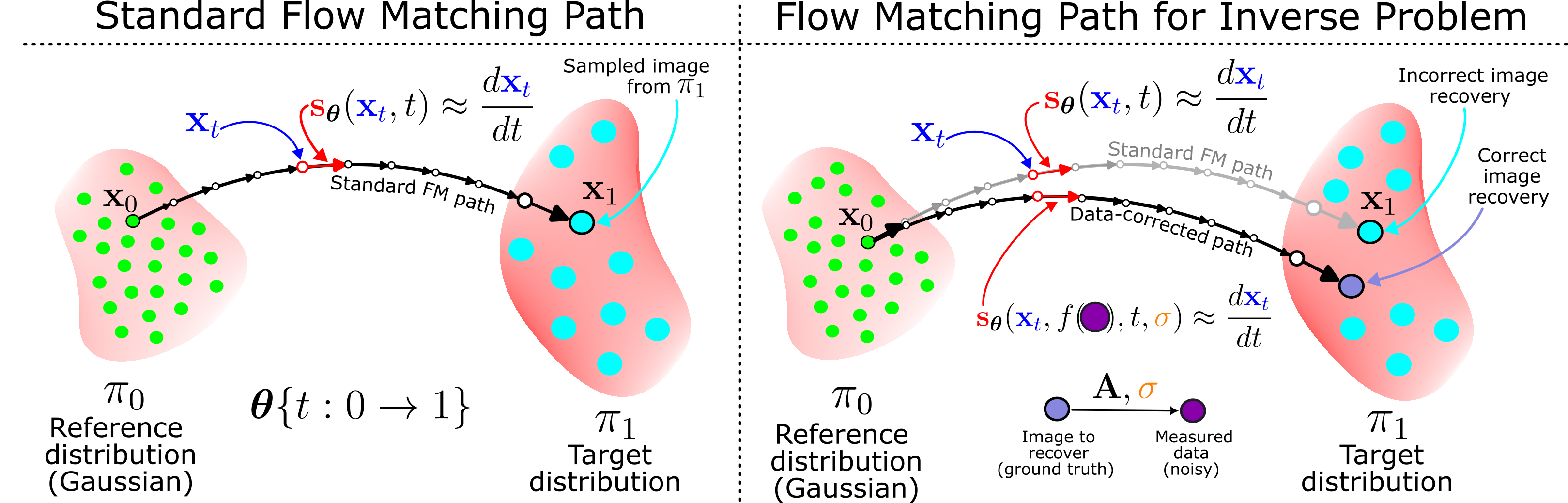}
    \caption{Schematics for standard flow matching (FM) (left) and flow matching for solving inverse problem characterized by the forward problem $\bfA$ along with an additive noise scale $\sigma$ (right). Here, $\bfs_\bftheta$ represents the trained network for estimating the velocity along the trajectory $\bfx_t$ at time $t$ between the reference distribution $\pi_0$ and the target distribution $\pi_1$, and $f$ represents a transformation on the measured (noisy) data which is used as data-embedding into the flow-matching network on the right.}
    \label{fig:FMschematic_without_with_data}
\end{figure}

In the canonical form of FM, the velocity is estimated from the interpolated vector $\bfx_t$. In the context of inverse problems, we also have a vector $\bfb$ (measured data) that contains additional information on $\bfx_1$ and therefore can be used to estimate the velocity towards the posterior. We now show that by doing a small change to the training process of FM, it is possible to solve inverse problems using the same concept.

To this end, notice that we train a velocity function $\bfs_{\bftheta}(\bfx_t, t)$ that takes in two arguments. In the context of a specific inverse problem, we have additional information for training the network, the measured data $\bfb$ and the noise level $\sigma$ in $\bfb$. Note that it is also possible to estimate the noise level directly from the data as proposed in \cite{Tenorio2011}. The data $\bfb$ can be used to point toward $\bfx_1$ even at time $t=0$, where $\bfx_t$ contains no information on $\bfx_1$ and can therefore improve the estimation of velocity. Moreover, the information about noise level $\sigma$ in the measured data during training also makes the estimator more robust to noise during inference.  
We thus propose to use the data and noise when estimating $\bfs_{\bftheta}$. To this end, we use a transformation $f$ of data $\bfb$ (to be discussed next), and let 
\begin{equation}
    \label{eq:mods}
    \bfs_{\bftheta} = \bfs_{\bftheta}(\bfx_t, f(\bfb), t, \sigma) = \bfs_{\bftheta}(\bfx_t, f(\bfA \bfx_1 + \sigma \bfz), t, \sigma),
\end{equation}
where $\bfz \sim \mathcal{N}(0, \bfI)$. To estimate $\bftheta$, we simply repeat the minimization process as before and match the flow where the data is a part of the estimated velocity, that is,
\begin{equation}
    \label{eq:velest_data}
    \hat \bftheta = {\rm arg}\min_{\bftheta} \mathcal{L}_1{(\bftheta)} =  {\rm arg}\min_{\bftheta} {\mathbb E}_{\bfx_0, \bfx_1, \sigma, t} \Big[ \left\| \bfs_{\theta}(\bfx_t, f(\bfA\bfx_1+\sigma\bfz), t, \sigma) + \bfx_0 - \bfx_1 \right\|^2 \Big], 
\end{equation}
where $\mathcal{L}_1$ represents the mean loss in the prediction of velocity. The trained network can then be used to invert new data. Let us assume that we are given some fixed vector $\bfb$ and we want to estimate $\bfx_1$. This can be done simply by solving the ODE
\begin{equation}
    \label{eq:inversion}
    {\frac {d\bfx_t}{dt}} = \bfs_{\bftheta}(\bfx_t, f(\bfb),  t, \sigma), \quad \bfx(0) = \bfx_0, \quad t\in [0,1],
\end{equation}
where $\bfb$ and $\sigma$ are now fixed.

As we show in our numerical experiments, having $\sigma$ as an input to the network plays an important role, generating an inversion methodology that is robust to different noise levels. While our method assumes that the noise standard deviation $\sigma$ from the data is known, this is not a strict limitation. In practice, $\sigma$ can often be estimated using classical image-adaptive heuristics such residual-based \cite{liu2006noise}, wavelet-domain estimators \cite{pimpalkhute2021digital}, or inferred by analyzing the eigenvalue distribution of patch covariance matrices \cite{chen2015efficient}.

An important question is the design of a network that integrates the information about $\bfb$ into the velocity estimation process. One important choice is the function $f$ that operates on $\bfb$. For many, if not most, inverse problems, the data $\bfb$ belongs to a different space than $\bfx$. Therefore, it is difficult to use this vector directly. The goal of the function $f$ is to transform $\bfb$ from the data space to the space of $\bfx$. One obvious approach to achieve this is to choose 
 \begin{equation}
 \label{eq:f}
 f(\bfb) = \bfA^{\top} \bfb.
 \end{equation}
This approach was used in \cite{mardani2018neural, adler2017solving}, and as shown in \Cref{sec:architecture}, can be successful for the transformation of the data into the image space. Other possible approaches can include fast estimation techniques for $\bfx$ given $\bfb$ such as the conjugate gradient least squares method \cite{hansen}. For the experiments presented here, we found that using the adjoint $\bfA^\top$ of the forward problem was sufficient.

To demonstrate these points, we consider the following toy example.

\begin{figure}
    \centering
    \begin{tabular}{cc}
    \includegraphics[scale=0.5]{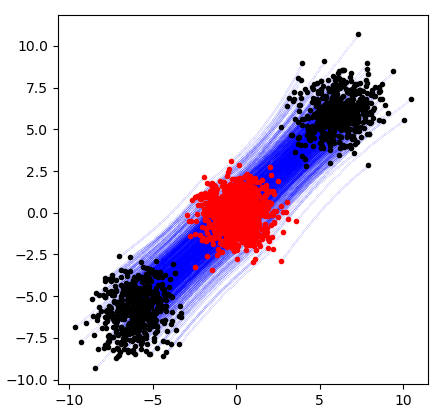} &
    \includegraphics[scale=0.5]{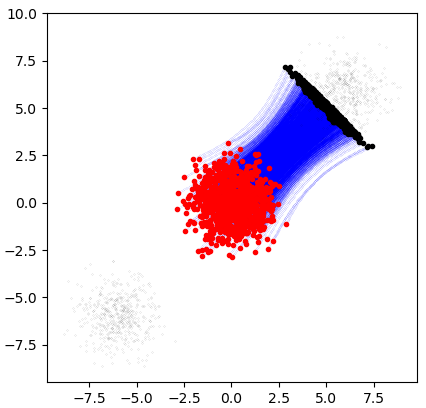} \\
    Generation without data & Generation with data 
    \end{tabular}
    \caption{Illustration of the Duathlon problem using flow matching (FM). Each $\bfx = [x_1, x_2]$ corresponds to bike and run times drawn from a prior distribution consisting of two Gaussians (shown in black). The red points denote observed measurements obtained through the forward model, $b = x_1 + x_2 + \epsilon$, and the blue curves show generated trajectories under FM. \textbf{Left:} FM generation without conditioning on data recovers the full mixture distribution, sampling across both Gaussian lobes. \textbf{Right:} FM generation with conditioning on measured data correctly identifies the appropriate lobe consistent with the observations and produces an ensemble of plausible solutions around it, thereby capturing the posterior variability of the inverse problem.}
    \label{fig:SIflow}
\end{figure}

\begin{example}
\label{ex:duathlon_problem}
{\bf The Duathlon problem:} We consider the duathlon problem where one records the total time it takes to perform a duathlon (bike and run). Given the total time, the goal is to recover the time it takes to perform each individual segment. The problem is clearly under-determined as one has to recover two number given a single data point.
Let $\bfx = [x_1, x_2]$ be the vector, where $x_1$ and $x_2$ represent the time it takes to finish the bike and run segments, respectively.
The data is simply, 
$$b = x_1 + x_2 + \epsilon.$$
Assume that we have a prior knowledge that the distribution of $\bfx$ is composed of two Gaussians. Using FM to generate data from these Gaussians is demonstrated in \Cref{fig:SIflow}.

The data is obtained by training a network, approximately solving the optimization problem in \Cref{eq:velest} and using \Cref{eq:velint} to integrate $\bfx_0$ that is randomly chosen from a Gaussian.

Now, in order to solve the inverse problem, we train a larger network that includes the data and approximately 
solves the optimization problem in \Cref{eq:velest_data}. Given the data $b$, we now integrate the ODE  \Cref{eq:inversion} to obtain a solution. The result of this integration is presented in \Cref{fig:SIflow} (right). We observe that not only did the process identify the correct lobe of the distribution, it also sampled many solutions around it. This enables us in obtaining not just a single but rather an ensemble of plausible solutions that can aid in exploring uncertainty in the result. 
\end{example}

\begin{figure}[ht]
\centering
\includegraphics[width=\textwidth, keepaspectratio]{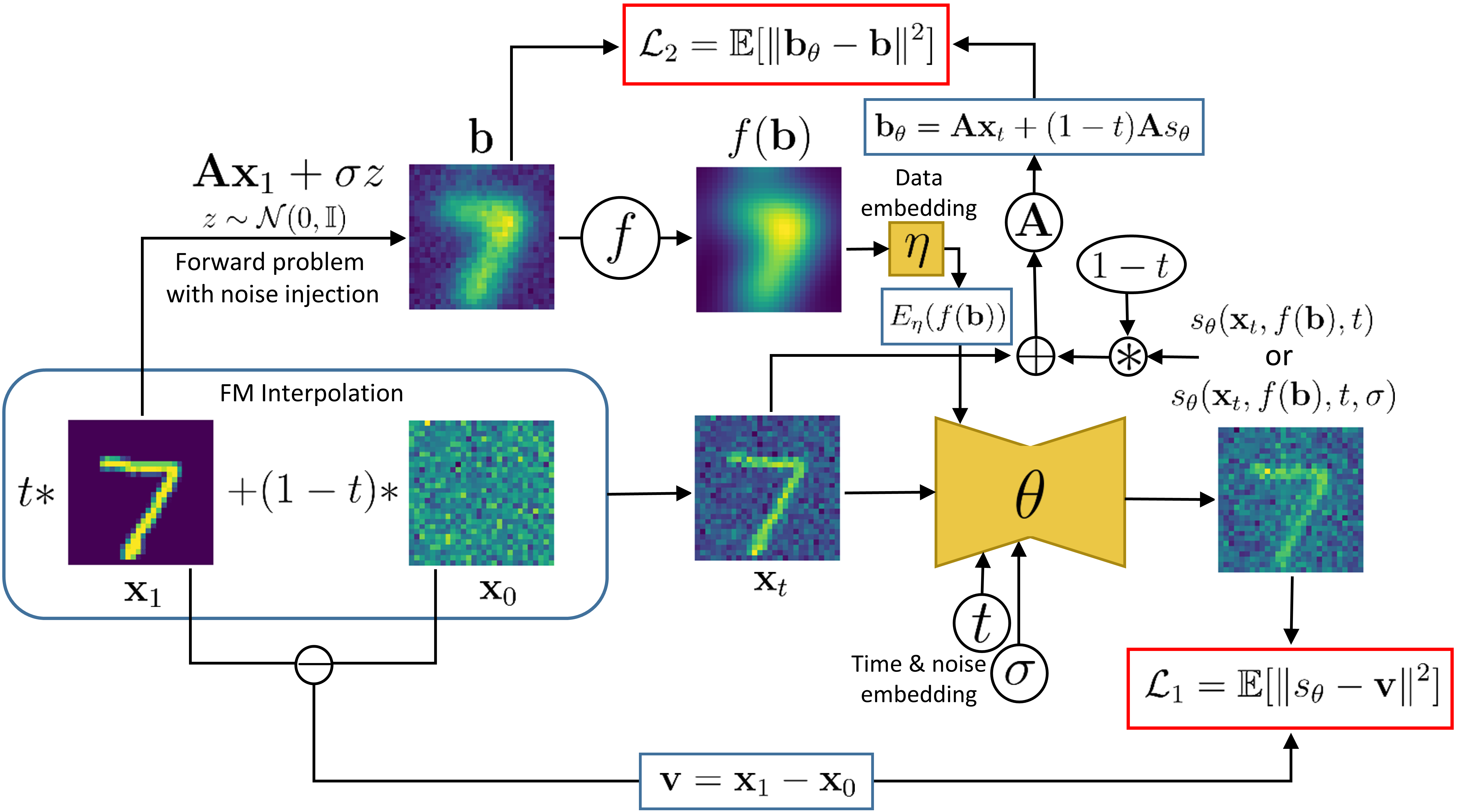}
    \caption{Schematic of the training process for the FM model for solving inverse problem, where the forward model is given by $\bfA$. This figure specifically represents the deblurring inverse problem. The network with parameters $\bftheta$ can may or may not incorporate the noise embedding (see \Cref{sec:numerical_experiments} for details). The two loss terms $\mathcal{L}_1$ and $\mathcal{L}_2$ represent the error in prediction of velocity  and misfit, respectively. In this figure, the transformation $f$ for generating the data embedding was chosen as $f = \bfA^\top$.}
    \label{fig:training_schematics}
\end{figure}

\subsection{Training the FM model for inverse problems}

When training the FM model, we solve the optimization problem given by \Cref{eq:velest}. In this process, one draws samples from $\bfx_0$ and $\bfx_1$, then randomly chooses $t\in[0,1]$ to generate the vector $\bfx_t$. 
In addition, we generate the data $\bfb$ by multiplying the matrix $\bfA$ with $\bfx_1$ and adding noise $\epsilon$ with a random standard deviation $\sigma$.

Next, one feeds the network $\bfx_t$, $t$, $\bfb$ and $\sigma$ to compute $\bfs_{\theta}(\bfx_t,\bfb, t, \sigma)$ and then compare it to the true velocity $\bfv = \bfx_1 - \bfx_0$.

While it is possible to train the network by comparing the computed velocity to its theoretical value, we have found that adding additional terms to the loss helps in getting a better estimate for the true velocity. In particular, when training the model for inverse problems, we add an additional term to the loss that relates to the particular inverse problem in mind. To this end, note that after estimating $\bfv$, we can estimate $\bfx_1$ using \Cref{eq:vfromxt}. Thus, we can estimate the data fit $\bfb_{\bftheta}$ for the estimated velocity, given by,
\begin{equation}
    \label{eq:datafit}
    \bfb_{\bftheta} = \bfA \bfx_t + (1-t) \bfA \bfs_{\bftheta}(\bfx_t,f(\bfb), t, \sigma).
\end{equation}
If the velocity is estimated exactly, then $\bfb_{\bftheta}=\bfb$. 

Note that if the velocity field $\bfs_\bftheta$ perfectly recovers the true conditional drift along the training trajectories and the measurement noise is zero (i.e., $\bfb = \bfA \bfx_1$), then by construction, the reconstruction $\hat{\bfx}_1 = \bfx_t + (1 - t) \bfs_\bftheta(\cdot)$ satisfies $\bfA \hat{\bfx}_1 = \bfb$, implying $\bfb_\theta = \bfb$. However, this identity holds strictly in the noiseless case. In the presence of measurement noise or when $\bfs_\bftheta$ is only an approximation of the true velocity field, $\bfb_\bftheta$ becomes a random variable that does not necessarily equal $\bfb$ for each sample. In this case, we can enforce data consistency in expectation by minimizing the discrepancy using $\| \bfA \hat{\bfx}_1 - \bfb \|^2$ through the $\ell_2$ loss. This ensures that the predicted measurements align with observed data in a statistical sense, even though exact sample-wise agreement is not guaranteed.

Therefore, a natural loss for the recovery is the comparison of the recovered data to the given data, meaning the the velocity function should honor the data $\bfb$, as well as the original distribution $\bfx_1$. We thus introduce the misfit loss term,
\begin{equation}
    \mathcal{L}_2(\bftheta) = {\mathbb E}_{\bfx_0, \bfx_1, \sigma, t} \Big[ \| \bfb_{\bftheta} - \bfb\|^2 \Big],
\end{equation}
that pushes the recovered $\bfv$ to generate an $\bfx_1$ that fits the measured data. 

To summarize, we modify the training by solving the following optimization problem,
\begin{equation}
    \label{eq:velestMod}
    \hat \bftheta =  {\rm arg}\min_{\bftheta} \Big\{{{\mathbb E}_{\bfx_0, \bfx_1, \sigma, t}  \| \bfs_{\theta}(\bfx_t, f(\bfb), t, \sigma) - \bfv \|^2} +\alpha {\mathbb E}_{\bfx_0, \bfx_1, \sigma, t}\|\bfb_{\bftheta} - \bfb\|^2\Big\},
\end{equation}

where $\alpha$ is a hyperparameter, which we set to 1. A schematic for training of FM model is given in \Cref{fig:training_schematics}.

It is important to note that while $\mathcal{L}_1$ is data-aware through its inputs ($f(\bfb)$), it does not by itself enforce that the image implied by the predicted velocity reproduces the measured data under the forward model. This is the main reason for introducing $\mathcal{L}_2$, which penalizes discrepancy between the observed data $\bfb$ and the data synthesized from the denoised, one-step estimate, $\hat{\bfx}_1$ passed through the forward operator, i.e., $\bfb_\theta = \bfA \hat{\bfx}_1$. In effect, $\mathcal{L}_2$ is a physics/data-consistency term that anchors the learned velocities so their implied reconstructions match the actual measurements, which is especially important in the early times $t \approx 0$  (when $\bfx_t$ carries little information about $\bfx_1$) and at higher noise levels. Practically, $\mathcal{L}_2$ behaves like a consistency regularizer on the velocity field (not a prior on the image), i.e., it does not constraint the null-space components of the forward model $\bfA$, but it suppresses solutions that fit $\mathcal{L}_1$ while violating the forward model in the range of $\bfA$. This complementary interplay between $\mathcal{L}_1$ (for amortized probability transport) and $\mathcal{L}_2$ (for explicit measured data fidelity) proved useful in our experiments.

\begin{figure}[h!]
    \centering
    \includegraphics[width=\linewidth]{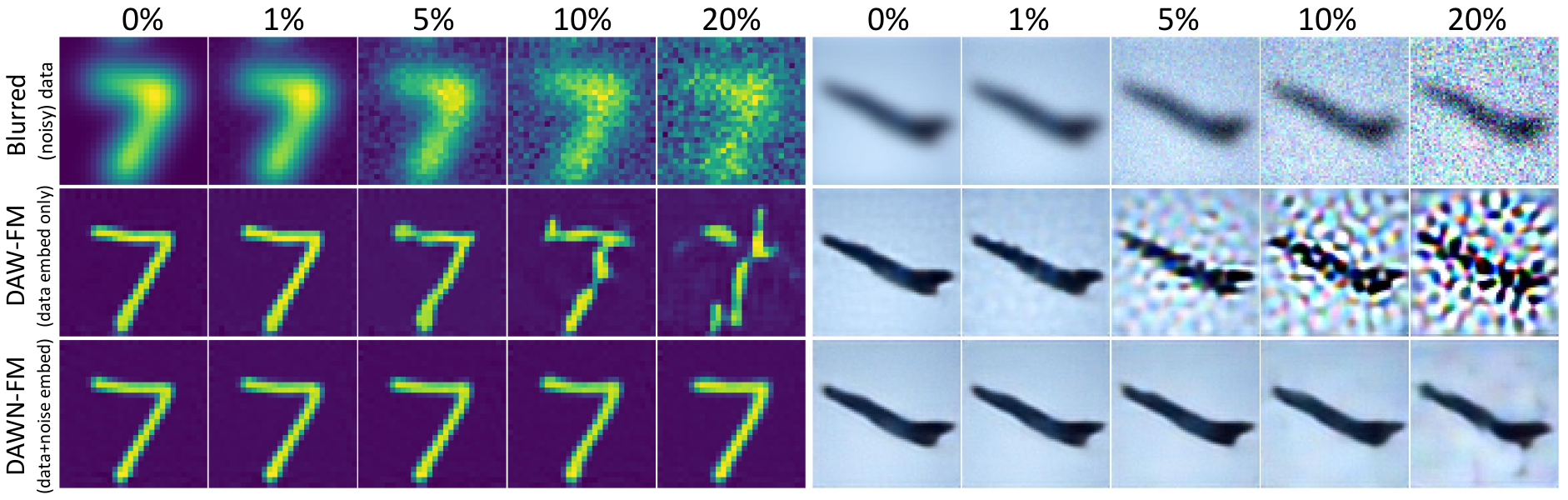}
    \caption{Example images from MNIST (left) and STL10 (right) datasets showing the recovery of deblurred images from the blurred (noisy) data (top panel) at different levels of noise (0\%, 1\%, 5\%, 10\%, 20\%). Our methods \sitd~ (middle panel) incorporates the embedding for the blurred data within the network, while \sitdn~(bottom panel) incorporates embedding for both data and noise levels. \sitdn~is superior to \sitd~at deblurred image recovery, especially at higher levels of noise (see \Cref{sec:numerical_experiments} for detailed implementation details for DAW-FM and DAWN-FM architectures.)}
    \label{fig:main_figure}
\end{figure}

\subsection{Architectures for data-aware and noise-informed velocity estimators}
\label{sec:architecture}

Our goal is to train a velocity estimator $\bfs_{\theta}(\bfx_t,\bfb, t, \sigma)$ and then compare it to the velocity $\bfv$. Our estimator is based on a UNet \cite{UNET2015} with particular embeddings for $\bfb, t$, and $\sigma$.

The embedding of time and noise is straightforward. Note that both $t$ and $\sigma$ are scalars. For their embedding, we use the method presented in \cite{croitoru2023diffusion}, which involves creating learnable embedding and adding them to the feature maps at each level of the network. 

A key component in making FM perform well for inverse problems is the careful design of a network that incorporates the data  within the training process. The resulting network can be thought of as a likelihood-free estimator \cite{thomas2022likelihood}, that estimates the velocity vector \(\bfv\), given the input vector $\bfx_t$, the time \(t\) and the noise level $\sigma$. As previously explained, we do not integrate $\bfb$ directly, but rather use $f(\bfb) = \bfA^{\top} \bfb$ and embed this vector in the network. 

The embedding of $\bfA^{\top}\bfb$ is performed
using a data encoder network. To this end, let 
\begin{equation}
\label{eq:data_enc}
E_{\bfeta}(\bfA^{\top}\bfb)    
\end{equation}
be an encoder network that is parameterized by $\bfeta$. The choice of using $f(\bfb) = \bfA^\top \bfb$ as the data embedding was primarily motivated by the need to map the measurement data back into the image domain, where the velocity estimation network operates. In our framework, the neural network is designed to predict the velocity field $s_\bftheta(\bfx_t, f(\bfb), t, \sigma)$ within the image space of $\bfx$, whereas the measured data $\bfb$ (e.g., sinograms in tomography) often resides in a different domain with a different dimensionality. Also, this design makes the network more robust and better conditioned, because the embedding $\bfA^\top \bfb$ acts as a non-local feature map. For the tomography problem, each pixel receives contributions from multiple rays (or projections), effectively aggregating contextual information from the entire measurement domain. This non-local property allows the network to exploit long-range dependencies and global structures in the reconstruction, rather than relying solely on local features.

Formally, the network estimates the velocity as,
\begin{equation}
    \label{eq:unetVel}
    \hat \bfv = \bfs_{\bftheta}(\bfx_t, E_{\bfeta}(\bfA^{\top}\bfb), t, \sigma).
\end{equation}
By embedding both time, data and noise vectors at each layer, the network can leverage additional information, leading to more accurate and robust predictions, especially for large noise levels.  

\subsection{Inference and uncertainty estimation}

\begin{figure}[h!]
    \centering
    \includegraphics[width=\linewidth]{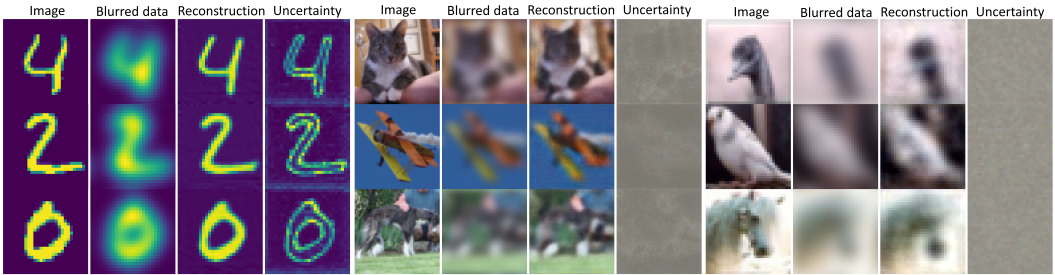}
    \caption{Computing uncertainty in the solutions obtained by FM for some example images from MNIST (left), STL10 (middle) and CIFAR10 (right) for the deblurring task. The reconstruction (posterior mean) is computed by averaging over runs from 32 randomly initialized $\bfx_0$. The uncertainty in the posterior mean is computed by evaluating its standard deviation for the predictions from 32 runs.}
    \label{fig:uncertainty}
\end{figure}

Our method is specifically aimed at highly ill-posed inverse problems, where regularization is often necessary to arrive at a stable solution. These problems often do not have a unique solution, and slight changes in input data can lead to large variations in the solution. Due to its stochastic nature, our method allows for the realization of multiple solutions since one can start with many random initial points $\bfx_0$ and evolve to multiple versions $\bfx_1$ of the solutions.
This property can be used to generate even better estimates and for the  estimation of uncertainty in the recovered images. 
To this end, assume that the ODE in \Cref{eq:inversion} is solved $M$ times starting each time at a different initial condition $\bfx_0^{(j)}$ and
let 
\begin{eqnarray}
\label{eq:realizations}
\bfx_1^{(j)} = \bfx_t(t=1, \bfx_0 = \bfx_0^{(j)}), \ \ j=1,\ldots, M 
\end{eqnarray}
 be the solution of the ODE in \Cref{eq:velest_data} obtained at time $t=1$ starting from point $\bfx_0^{(j)}$ at $t=0$. The points $\bfx_1^{(j)}$ for $j=1,\ldots,M$ represent an ensemble of solutions, that are sampled from the posterior $\pi(\bfx_1|\bfb)$. Given these points, it is simple to estimate the mean and standard deviation of the posterior distribution. In particular,  we define
\begin{equation}
    \label{meanstd}
    {\bar \bfx}_1 = {\frac 1M} \sum_{j=1}^M \bfx_1^{(j)}, \quad {\sigma}_{{\bar \bfx}_1} = {\frac 1M} \sum_{j=1}^{M} \|\bfx_1^{(j)} - {\bar \bfx}_1 \|^2,  
\end{equation}
as the mean and standard deviation of the estimated solutions.
In particular, ${\bar \bfx}_1$ approximates the posterior mean and ${\sigma}_{{\bar \bfx}_1}$ estimates its standard deviation. It is well known that, for many problems the posterior mean can have a lower risk compared to other estimators (e.g. the Maximum A-Posteriori estimator). For an elaborate discussion on the properties of such estimators, see \cite{Tenorio2011, somersallo, CalvettiSomersalo2005}. While such estimators are typically avoided due to computational complexity, the computational framework presented here allows us to compute them relatively easily.

An example for this process is shown in \Cref{fig:uncertainty}. Here,  the solutions over multiple runs were averaged to generate an effective reconstruction of the original image (the posterior mean). Moreover, to quantify uncertainty in the reconstruction process, we computed standard deviation over the solutions at each pixel from multiple runs. In the uncertainty maps for MNIST, the uncertainty is concentrated along the edges of the digits. This occurs because the FM model introduces slight variations in how it reconstructs the boundary between the digit and the background. Since this boundary is sharp, any slight differences in how this edge is defined in different reconstructions lead to higher uncertainty along the edges. On the other hand, for STL10 and CIFAR10 images, the boundary between objects and background is often less distinct. The background might contain detailed textures or noise that blends into the object, making it harder for the model to distinguish clear boundaries. Hence, the uncertainty maps for these datasets do not exhibit the same clear edge-focused uncertainty as in MNIST. The lack of a clear boundary means that the reconstruction's variability spreads more evenly across the entire image.

\section{Details about the Inverse Problems}
\label{sec:inverse_problems_details}
\subsection{Details for image deblurring}
\label{app:image_deblurring}
In this section, we provide technical details about the image deblurring inverse problem used in this work. For this inverse problem, the forward operator $\bfA$ (a blurring operator) can be represented as a convolution of the input image $I(x, y)$ with a Gaussian Point Spread Function (PSF) kernel, where the kernel is characterized by a standard deviation. The blurring process can be mathematically described as a convolution between a sharp image $I$ and a (Gaussian) blurring kernel $K$, where $I_{\text{blur}}$ is the resultant blurred image,
\begin{equation}
I_{\text{blur}}(x, y) = \bfA I(x, y) + \epsilon (x, y) = I (x, y) \ast K(x, y) + \epsilon (x, y),
\end{equation}
where $I_{\text{blur}}(x, y)$ is the blurred image at pixel location $(x, y)$, $I(x, y)$ is the original sharp image, $K(x,y)$ is the PSF, and $\epsilon(x, y)$ is the additive noise, which is also assumed to be Gaussian with zero mean and a standard deviation equal to $p\%$ of the range of $I (x, y) \ast K(x, y)$, where $p$ was chosen uniformly between 0\% and 20\% during training. The Gaussian PSF kernel is given by,
\begin{equation}
    K(x, y)= \frac{1}{2 \pi \sigma_x \sigma_y} \exp\bigg(-\frac{x^2}{\sigma_x^2} -\frac{y^2}{\sigma_y^2} \bigg).
\end{equation}
All our experiments were run with $\sigma_x = \sigma_y = 3$. The goal of image deblurring is to recover the original sharp image \( I(x, y) \) from the blurred image \( I_{\text{blur}}(x, y) \). This is typically ill-posed due to the possible presence of noise \( \eta(x, y) \) and the loss of high-frequency information caused by the blur. 

The convolution operation can be described as a multiplication in the Fourier domain,
\begin{equation}
    \Tilde{B}(u, v) = \Tilde{I}(u, v) \cdot \Tilde{K}(u, v),
\end{equation}
where \(\Tilde{I}(u,v) = \mathcal{F}\{I(x, y)\} \), \(\Tilde{K}(u, v) = \mathcal{F}\{K(x, y)\} \) are the Fourier transforms of the sharp image, and the PSF, respectively and \( \mathcal{F} \) represents the Fourier transform. The blurred image can then be obtained using the inverse Fourier transform,
\begin{equation}
     I_{\text{blur}}(x, y) =  \mathcal{F}^{-1} \{\Tilde{B}(u, v)\}.
\end{equation}
Moreover, $\bfA^\top$, the adjoint of operator $\bfA$, which is used to compute the data-embedding $\bfA^\top \bfb$ for the network, for some data $\bfb$ is easy to obtain. Since the image blurring operator is symmetric and self-adjoint, we have $\bfA = \bfA^\top$. In this work, we compare our method to diffusion-based method \cite{chung2022diffusion} and InverseUNetODE \cite{eliasof2024completedeeplearningmethod} to highlight the robustness of our method against competing baselines (see \Cref{sec:numerical_experiments}).

\subsection{Details for tomography}
\label{app:tomography}
In tomography, the goal is to reconstruct the internal image of an object from a series of projections (sinograms) taken at various angles. Mathematically, this problem is formulated as solving an inverse problem: the original image $I$ is mapped to its projections $S$ through the Radon transform $\bfA$, and reconstruction consists of estimating $I$ from these measured projections.


In this work, we consider the standard parallel-beam tomography model, in which the forward problem can be described as,
\begin{equation}
\text{vec}({S}) = \bfA \text{vec}(I_{\text{pad}}) + \epsilon,
\end{equation}
where, $\text{vec}(\cdot)$ represents vectorization (flattening), \( \text{vec}(S) \in \mathbb{R}^{N_{\text{angles}} \cdot N_{\text{detectors}}} \) is the flattened sinogram (the set of projections), \( \text{vec}(I_{\text{pad}}) \in \mathbb{R}^{N_{\text{pixels}}} \) is the padded and flattened image of dimension \( N_{\text{pixels}}\), \( \bfA \in \mathbb{R}^{(N_{\text{angles}} \cdot N_{\text{detectors}}) \times N_{\text{pixels}}} \) is the tomography projection matrix which describes the interaction of each ray with every pixel, and $N_\text{angles}$, $N_\text{detectors}$, and $N_\text{pixels}$ represent the number of tomographic projection angles, detectors, and pixels (in the padded image), respectively. The 2D sinogram $S \in \mathbb{R}^{N_{\text{angles}} \times  N_{\text{detectors}}}$ is then obtained by reshaping $\text{vec}(S)$. The term $\epsilon$ is the additive noise to the data, which was chosen as $p\%$ of the range of values in $\bfA \text{vec}(I_\text{pad})$, with $p$ lying between 0\% and 20\% during training. For our experiments, we set \(N_\text{angles} = 180\), and $N_\text{detectors} = 2s + 1$, where $s \times s$ is the dimension of the original unpadded images. 

The image is first padded to account for the fact that projections are taken beyond the boundaries of the object in the image. Using a zero padding of size $s/2$ on all four sides of the image, the padded version is an image of dimension $2s \times 2s$. The matrix $\bfA$ is multiplied with the flattened version of the padded image generating the sinogram data.   

A simple approach to reconstruct the image from measured sinograms is backprojection, which uses the adjoint operator $\bfA^\top$ of the forward projection matrix. The reconstructed image is then,
\begin{equation}
\text{vec}(I_{\text{pad}}) = \bfA^\top \text{vec}(S).
\end{equation}
The resulting $\text{vec}(I_\text{pad})$ is reshaped and cropped back to size $s \times s$. However, because $\bfA^\top$ is not the analytic inverse of $\bfA$, the reconstruction is typically blurred. This distinction is fundamental: for tomography, $\bfA^\dagger \neq \bfA^\top$, where $\bfA^{\dagger}$ represents the Moore-Penrose pseudoinverse of the Radon transform.

\textbf{Filtered Backprojection (FBP):} To address the blurring inherent in plain backprojection, classical tomography uses Filtered Backprojection (FBP) - an analytic inversion technique that restores high-frequency information lost during the projection process. FBP involves two main steps:
\begin{itemize}
    \item Filtering: Each 1D projection $S(\phi,z)$ (for angle $\phi$ and detector coordinate $z$) is Fourier transformed and multiplied by a frequency-domain filter $|\omega|$. Commonly used filters include ramp and Shepp-Logan filters. This step compensates for the attenuation of high frequencies in the measurement process.
    \item Backprojection: The filtered projections are then inverse Fourier transformed and backprojected into the image space over all projection angles.
\end{itemize}
Mathematically, the reconstruction can be written as, 
\begin{equation}
 I(x,y) = \int_{0}^{\pi} \bigg[\int_{-\infty}^{\infty} \Tilde{S}(\phi, \omega) |\omega| e^{2\pi i \omega z} d\omega \bigg] d\phi,
\end{equation}
where $\Tilde{S}(\phi, \omega)$ is the Fourier transform of $S(\theta, z)$ with respect to $z = x \text{cos} \phi + y \text{sin} \phi$ and $\omega$ represents the ramp filter. This additional filtering step is what distinguishes $\bfA^{\dagger}$ from $\bfA^\top$: while $\bfA^\top$ simply redistributes projection values back into the image domain, FBP explicitly corrects for the frequency imbalance to yield a sharper reconstruction.

For this task, we used $\bfA^\top \bfb$ instead of the analytic inverse to inject data $\bfb$ into the network because (i) $\bfA^\top$ provides a simple way to inject measurement information into the image domain, ensuring that the reconstructed estimate is consistent with the geometry of the data (to avoid training instabilities arising from different-sized multi-channel inputs), and (ii) our neural flow-matching approach learns to refine the backprojected image and compensate for the blur that would otherwise arise, effectively learning a data-driven analogue of the analytic filter step in FBP. In other words, our method treats $\bfA^\top \text{vec}(S)$ as a physically consistent but blurred prior, which the neural network then refines into a high-fidelity reconstruction. This hybrid approach leverages the interpretability and geometric fidelity of classical reconstruction while exploiting the representational power of deep generative models to achieve superior reconstruction quality. In this work, we compare our method against the FBP baseline with the ramp filter (see \Cref{sec:numerical_experiments}).

\section{Numerical Experiments}
\label{sec:numerical_experiments}
In this section, we experiment with our method on a few common datasets and two broadly applicable inverse problems: image deblurring and tomography (see \Cref{sec:inverse_problems_details} for details). We provide additional information on the experimental settings, hyperparameter choices, and network architectures in \Cref{tab:expt_details_deblurring_diffusion,tab:expt_details_deblurring_inverseunetode,tab:expt_details_deblurring_ourmethods} for image deblurring task and \Cref{tab:expt_details_tomography_ourmethods} for tomography task in \Cref{app:experiment_setting}. 

\textbf{Training methodology.} For our experiments, we considered two types of data-aware velocity estimator networks: (i) \sitd : the estimator with no noise-embedding, and (ii) \sitdn~: the estimator with trainable noise-embedding. We employed antithetic sampling during training. Starting with an input batch of clean images $\bfx_1^\prime$, an $\bfx_0^\prime \sim \mathcal{N}(0, \bfI)$ was sampled and antithetic pairs $\bfx_1$ and $\bfx_0$,
\begin{equation}
\bfx_1 = \begin{bmatrix} \bfx_1^\prime \\ \bfx_1^\prime\end{bmatrix}, \quad \bfx_0 = \begin{bmatrix} \bfx_0^\prime \\ -\bfx_0^\prime\end{bmatrix},
\end{equation}
were generated by concatenation along the batch dimension. For a large number of samples, the sample mean of independent random variables converges to the true mean. However, the convergence can be slow due to the variance of the estimator being large. Antithetic sampling helped reduce this variance by generating pairs of negatively correlated samples, thereby improving convergence. The data was generated by employing the forward model along with a Gaussian noise injection, $\bfb = \bfA \bfx_1 + \sigma \bfz$, where $\bfz \sim \mathcal{N}(0, \bfI)$. The value of $\sigma$ was set to $p\%$ of the range of values in the data $\bfb$, where $p$ was sampled uniformly in (0, 20). $\bfx_t$ and $\bfv$ were computed from $\bfx_1$ and $\bfx_0$ following \Cref{eq:traj} and \Cref{eq:vel}. 

\textbf{Network implementation.} To represent the temporal dynamics of flow matching and the effect of measurement noise within a unified network, we embed both the interpolation time $t$ and the noise level $\sigma$ into learnable feature vectors that modulate the network at every resolution. scalar time input $t \in [0,1]$ is passed through a linear projection to produce a time embedding, which is then broadcast and added as a per-channel bias at each of the encoder, bottleneck, and decoder blocks. This allows the network to adapt its feature transformations depending on the location on the flow trajectory. Similarly, the scalar noise level $\sigma$ is embedded into a noise feature vector. At each resolution, this vector is reshaped and combined multiplicatively with the backprojected data $\bfA^\top \bfb$, producing a noise-aware, spatially structured modulation of the features. In practice, this mechanism enables the model to rely more strongly on data consistency when the noise is low, while leaning more on the learned prior when the noise level is high. Together, the time embedding provides trajectory awareness and the noise embedding provides data-aware regularization, yielding a network that is both flow- and noise-adaptive.
 
To incorporate information from the measured data into the network at every resolution, we introduce a dedicated data embedding pathway that operates on the backprojected measurements $\bfA^\top \bfb$. At each encoder and decoder level, $\bfA^\top \bfb$ is first rescaled to match the current spatial resolution of the feature maps. This rescaled tensor is then processed by a lightweight convolutional module consisting of two consecutive $3 \times 3$ convolutions with SiLU activations (overall denoted as $E_\bfeta(\cdot)$ in \Cref{eq:data_enc}). This produces a set of feature maps that encode the structure of the backprojected measurements in a form that can be directly combined with the learned network features. By injecting these data features additively into each block, the network is explicitly encouraged to maintain coherence with the measured data throughout the forward and backward passes. In combination with the noise-conditioned modulation described above, the data embedding serves as the primary mechanism through which the measured observations influence the learned velocity field. Note that using the transformation $E_\bfeta (f(\bfb)) = E_\bfeta (\bfA^\top \bfb$), the predicted velocity of the estimator was $s_\bftheta(\bfx_t,  E_\bfeta(\bfA^\top\bfb), t)$ for the \sitd~estimator (no noise embedding) and $s_\bftheta(\bfx_t, E_\bfeta(\bfA^\top\bfb), t, \sigma)$ for the \sitdn~estimator. The loss for an epoch was computed as given in \Cref{eq:velestMod}.

\textbf{Inference}. For inference on a noisy data, $\bfb = \bfA \bfx_1 + \sigma \bfz$ for some image $\bfx_1$, we start with a randomly sampled $\bfx_0 \sim \mathcal{N}(0, \bfI)$ at $t=0$ and perform Fourth-Order Runga Kutta numerical integration to solve the ODE in \Cref{eq:inversion} with a step size $h = 1/100$, where the velocity was computed using the trained estimator as $s_\bftheta(\bfx_t, E_\bfeta(\bfA^\top \bfb), t)$ for the \sitd~estimator and $s_\bftheta(\bfx_t, E_\bfeta(\bfA^\top \bfb), t, \sigma)$ for the \sitdn~estimator at time $t$. For each image $\bfx_1$, the inference was run 32 times starting from 32 different realizations of $\bfx_0$ and the resulting reconstructed images (which are samples from the posterior $\pi (\bfx_1 | \bfb)$) were averaged to generate the final recovered image. The uncertainty at the pixel-level could also be estimated from these samples by computing their standard deviation, as shown in \Cref{fig:uncertainty}.  

\textbf{Baselines.} For the image deblurring task, we compare our methods \sitd~and \sitdn~to diffusion-based image deblurring \cite{chung2022diffusion} and InverseUNetODE \cite{eliasof2024completedeeplearningmethod}. The details for the training setup and hyperparameter choices for the baselines have been provided in \Cref{app:experiment_setting}. All baselines were trained with Gaussian noise injection in the data in the same way as used for training of \sitd~and \sitdn~methods for fair comparison.  

\textbf{Metrics.} For performance evaluation, we compute mean squared error (MSE), misfit, structural similarity index measure (SSIM) and peak signal-to-noise ratio (PSNR) between the ground truth and the reconstructed image. Additional details on these metrics are provided in \Cref{app:evaluation_metrics}.

\subsection{Image deblurring}
\label{subsec:image_deblurring}
\begin{figure}[]
    \centering
    \includegraphics[width=\linewidth]{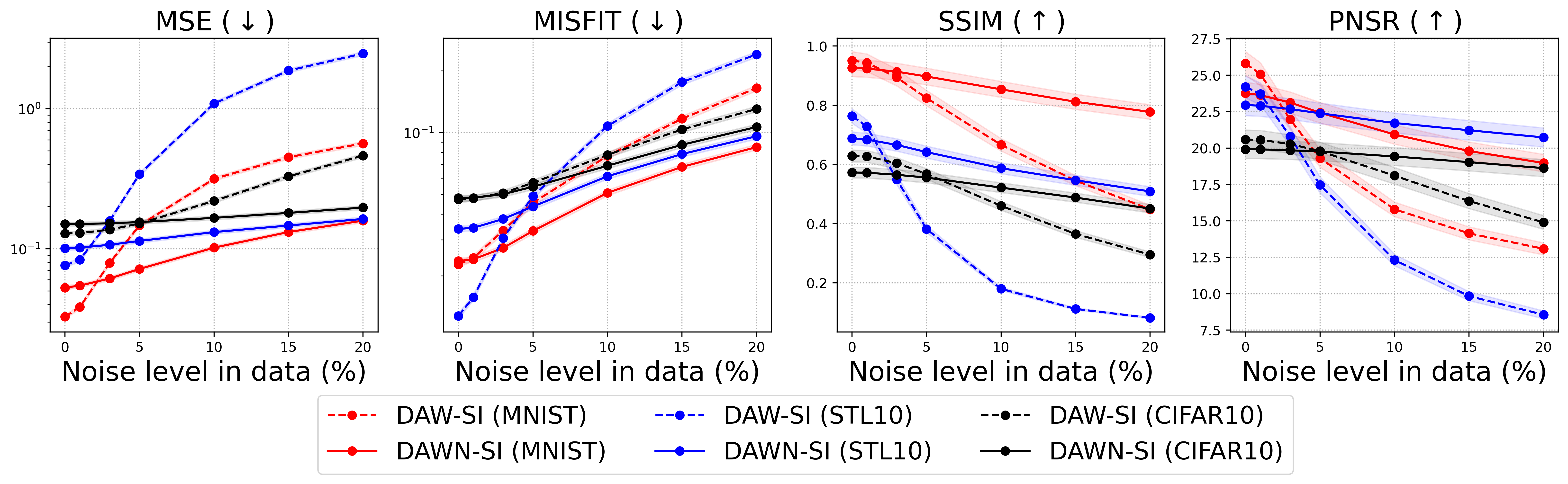}
    \caption{Assessing the sensitivity of the metrics for the recovered deblurred image as a function of the level of noise ($p\%$) added to the data. The \sitdn~is more robust to noise across all three datasets, MNIST, STL10 and CIFAR10 than \sitd, with the crossing point happening at values of $p$ $\lesssim$ $5\%$ for all metrics.}
    \label{fig:noise_ablation_deblurring}
\end{figure}

\begin{figure}[h]
    \centering
    \includegraphics[trim=40 0 40 0, clip, width=0.8\linewidth]{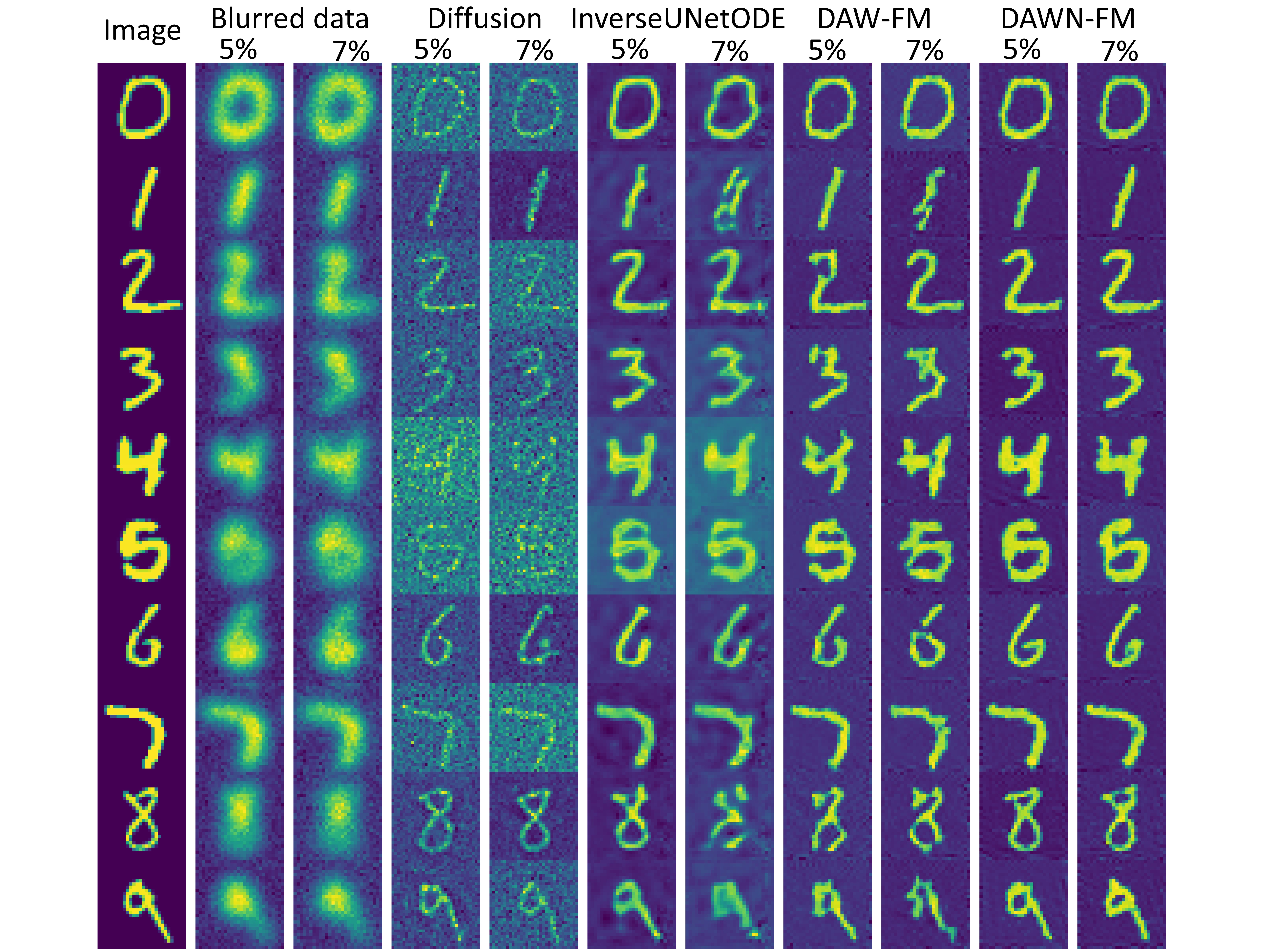}
    \caption{Comparison between different methods for the \textbf{image deblurring} task on the \textbf{MNIST} dataset for noise levels $p = 5\%$ and 7\% added to the blurred data.}
    \label{fig:AppendixImages_mnist}
\end{figure}

\begin{figure}[h]
    \centering
    \includegraphics[width=0.9\linewidth]{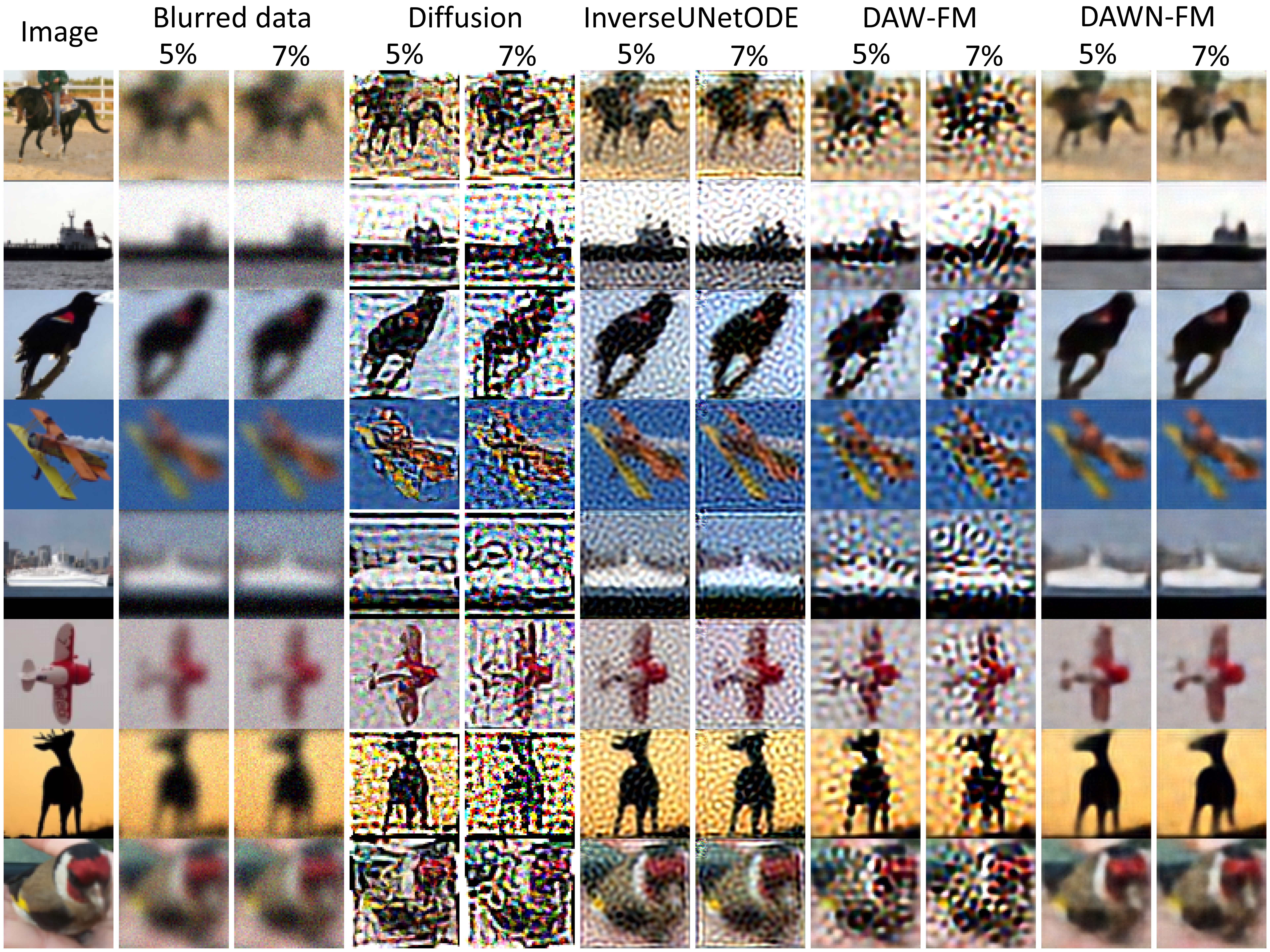}
    \caption{Comparison between different methods for the \textbf{image deblurring} task on the \textbf{STL10} dataset for noise levels $p = 5\%$ and 7\% added to the blurred data.}
    \label{fig:AppendixImages_stl10}
\end{figure}

\begin{figure}[h]
    \centering
    \includegraphics[width=0.9\linewidth]{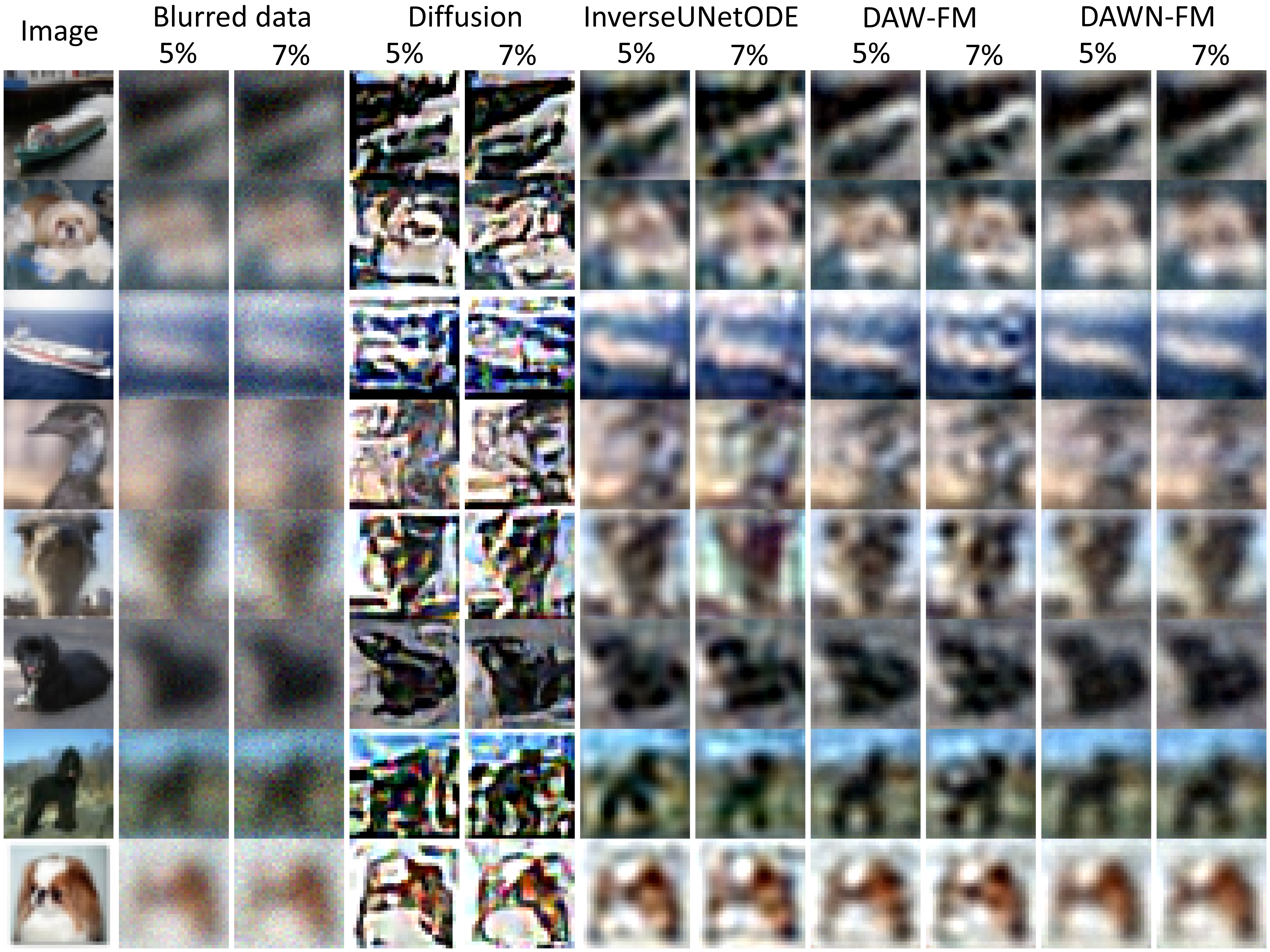}
    \caption{Comparison between different methods for the \textbf{image deblurring} task on the \textbf{CIFAR10} dataset for noise levels $p = 5\%$ and 7\% added to the blurred data.}
    \label{fig:AppendixImages_cifar10}
\end{figure}

We present the results of image deblurring task in \Cref{tab:deblurring_results} at noise level $p = 5\%$ in blurred data for MNIST, STL10 and CIFAR10 datasets. For all metrics, \sitdn~beats all other methods for all datasets by large margins, except for the CIFAR10 dataset, where \sitd~performed marginally better than \sitdn. Moreover, on the CIFAR10 dataset, the diffusion model fits the data the best achieving the lowest value of the misfit metric. We also performed ablation studies for different noise levels in data for both \sitd~and \sitdn~models, as shown in \Cref{fig:noise_ablation_deblurring}. For all metrics, \sitdn~was more robust to noise in data, especially for higher noise levels over all datasets with the crossing point happening at values of $p \lesssim 5\%$ for all metrics. Some example images for the deblurring task for different levels of noise in data are presented in \Cref{fig:AppendixImages_mnist,fig:AppendixImages_stl10,fig:AppendixImages_cifar10} in \Cref{app:additional_examples}. Based on empirical evidence from training, we believe that the results for both \sitd~and \sitdn~could be improved further with more training epochs.  

\begin{table}[]
\centering
\caption{Comparison between the performances of different methods (Diffusion, InverseUNetODE, and our methods \sitd~and \sitdn) for the deblurring task for MNIST, STL10 and CIFAR10 datasets using metrics MSE, MISFIT, SSIM and PSNR. All the evaluations have been presented on the test set for a large blurring kernel (for details, see \Cref{app:image_deblurring}) at a noise level $p = 5\%$ in the blurred data. Here, we report mean $\pm$ standard deviation over all images in the test set. First and second best performances are shown in \textbf{bold} and \textcolor{red}{red}, respectively.}
\label{tab:deblurring_results}
\resizebox{\columnwidth}{!}{%
\begin{tabular}{@{}clcccc@{}}
\toprule
\multirow{2}{*}{\textbf{Dataset}} &
  \multicolumn{1}{c}{\multirow{2}{*}{\textbf{Methods}}} &
  \multirow{2}{*}{\textbf{MSE ($\downarrow$)}} &
  \multirow{2}{*}{\textbf{MISFIT ($\downarrow$)}} &
  \multirow{2}{*}{\textbf{SSIM ($\uparrow$)}} &
  \multirow{2}{*}{\textbf{PSNR ($\uparrow$)}} \\
                         & \multicolumn{1}{c}{} &               &               &               &                \\ \midrule
\multirow{4}{*}{MNIST}   & Diffusion            & 6.078 $\pm$ 8.508 & 0.045 $\pm$ 0.008 & 0.248 $\pm$ 0.178 & 6.538 $\pm$ 6.980  \\
                         & InverseUNetODE       & 0.167 $\pm$ 0.072 & 0.055 $\pm$ 0.012 & 0.714 $\pm$ 0.090 & 18.394 $\pm$ 1.946 \\
                         & \sitd~(Ours)         & \textcolor{red}{0.156 $\pm$ 0.078} & \textcolor{red}{0.043 $\pm$ 0.012} & \textcolor{red}{0.825 $\pm$ 0.062} & \textcolor{red}{18.899 $\pm$ 2.472} \\
                         & \sitdn~(Ours)        & \textbf{0.073 $\pm$ 0.041} & \textbf{0.032 $\pm$ 0.009} & \textbf{0.901 $\pm$ 0.042} & \textbf{22.277 $\pm$ 2.612} \\ \midrule
\multirow{4}{*}{STL10}   & Diffusion            & 3.712 $\pm$ 3.900 & 0.050 $\pm$ 0.012 & 0.170 $\pm$ 0.057 & 8.581 $\pm$ 3.765  \\
                         & InverseUNetODE       & \textcolor{red}{0.299 $\pm$ 0.110} & 0.084 $\pm$ 0.063 & 0.360 $\pm$ 0.059 & \textcolor{red}{18.427 $\pm$ 1.342} \\
                         & \sitd~(Ours)         & 0.344 $\pm$ 0.132 & \textcolor{red}{0.049 $\pm$ 0.012} & \textcolor{red}{0.382 $\pm$ 0.064} & 17.434 $\pm$ 1.465 \\
                         & \sitdn~(Ours)        & \textbf{0.113 $\pm$ 0.056} & \textbf{0.043 $\pm$ 0.013} & \textbf{0.644 $\pm$ 0.082} & \textbf{22.423 $\pm$ 1.904} \\ \midrule
\multirow{4}{*}{CIFAR10} & Diffusion            & 1.049 $\pm$ 0.749 & \textbf{0.042 $\pm$ 0.012} & 0.272 $\pm$ 0.081 & 12.104 $\pm$ 2.888 \\
                         & InverseUNetODE       & 0.168 $\pm$ 0.080 & 0.067 $\pm$ 0.027 & 0.544 $\pm$ 0.075 & 19.388 $\pm$ 1.587 \\
                         & \sitd~(Ours)         & \textbf{0.152 $\pm$ 0.071} & 0.057 $\pm$ 0.025 & \textbf{0.567 $\pm$ 0.077} & \textbf{19.765 $\pm$ 1.520} \\
                         & \sitdn~(Ours)        & \textcolor{red}{0.156 $\pm$ 0.077} & \textcolor{red}{0.055 $\pm$ 0.023} & \textcolor{red}{0.554 $\pm$ 0.081} & \textcolor{red}{19.743 $\pm$ 1.712} \\ \bottomrule
\end{tabular}%
}
\end{table}

\begin{figure}[]
    \centering
    \includegraphics[trim=0 170 0 150, clip, width=\linewidth]{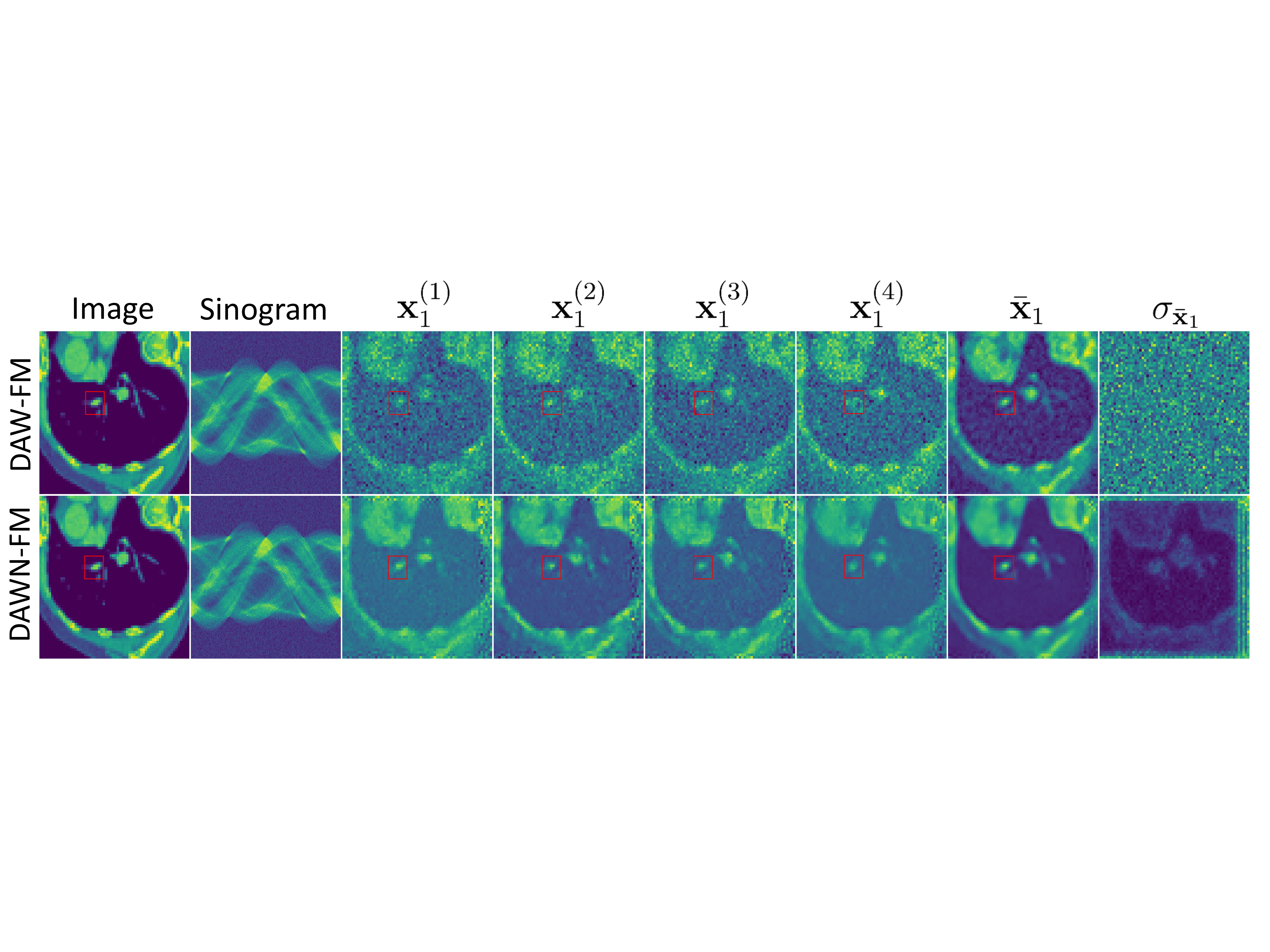}
    \caption{An example from OrganAMNIST dataset showing the different samples (labeled $\bfx^{(1)}_1$ to  $\bfx^{(4)}_1$) drawn from the posterior starting from different initial $\bfx_0$ for both \sitd~(top) and \sitdn~(bottom). The posterior mean $\bar{\bfx}_1$ and standard deviation ${\sigma}_{{\bar \bfx}_1}$ were computed from 32 such samples, given the sinogram data (at noise level $p = 5\%$). The red box highlights a lobe, which exhibits variability in its size and shape across different samples. This variability reflects the model's uncertainty in reconstructing (the boundary of) such anatomical features.}
    \label{fig:medmnist_uncertainty}
\end{figure}

\subsection{Tomography}

\begin{table}[]
\centering
\caption{Comparison between the performances of our methods (\sitd~and \sitdn) for the tomography task for OrganAMNIST and OrganCMNIST datasets using metrics MSE, MISFIT, SSIM and PSNR. All the evaluations have been presented on the test set for $N_\text{angles} = 180$, and $N_\text{detectors} = 2s+1$, where $s \times s$ is the image dimension (for details, see \Cref{app:tomography}) at a noise level $p = 5\%$ in the sinogram data. Here, we report mean $\pm$ standard deviation over 5000 images in the test set. First and second best performances are shown in \textbf{bold} and \textcolor{red}{red}, respectively.}
\label{tab:tomography_results}
\resizebox{\columnwidth}{!}{%
\begin{tabular}{@{}clcccc@{}}
\toprule
\multirow{2}{*}{\textbf{Dataset}} &
  \multicolumn{1}{c}{\multirow{2}{*}{\textbf{Methods}}} &
  \multirow{2}{*}{\textbf{MSE ($\downarrow$)}} &
  \multirow{2}{*}{\textbf{MISFIT ($\downarrow$)}} &
  \multirow{2}{*}{\textbf{SSIM ($\uparrow$)}} &
  \multirow{2}{*}{\textbf{PSNR ($\uparrow$)}} \\
                             & \multicolumn{1}{c}{} &               &               &               &                \\ \midrule
\multirow{2}{*}{OrganAMNIST} & FBP (ramp filter)     & 0.014 $\pm$ 0.006 & \textbf{0.029 $\pm$ 0.006} & 0.469 $\pm$ 0.114  & 18.047 $\pm$ 2.129 \\ 
                            & \sitd~(Ours)               & \textcolor{red}{0.008 $\pm$ 0.002} & 0.041 $\pm$ 0.051 & \textcolor{red}{0.575 $\pm$ 0.108} & \textcolor{red}{20.389 $\pm$ 1.767} \\
                             & \sitdn~(Ours)               & \textbf{0.004 $\pm$ 0.001} & \textcolor{red}{0.036 $\pm$ 0.060} & \textbf{0.713 $\pm$ 0.090} & \textbf{23.244 $\pm$ 1.553} \\ \midrule
\multirow{3}{*}{OrganCMNIST} & FBP (ramp filter)     & 0.015 $\pm$ 0.006 & \textbf{0.029 $\pm$ 0.008} & 0.437 $\pm$ 0.128 & 17.413 $\pm$ 2.271 \\ 
                            & \sitd~(Ours)                & \textcolor{red}{0.014 $\pm$ 0.008} & 0.156 $\pm$ 1.424 & \textcolor{red}{0.511 $\pm$ 0.140} & \textcolor{red}{17.881 $\pm$ 2.624}` \\
                             & \sitdn~(Ours)              & \textbf{0.007 $\pm$ 0.007} & \textcolor{red}{0.151 $\pm$ 1.287} & \textbf{0.675 $\pm$ 0.116} & \textbf{21.630 $\pm$ 3.199} \\ \bottomrule
\end{tabular}%
}
\end{table}

\begin{figure}[h]
    \centering
    \includegraphics[width=0.9\linewidth]{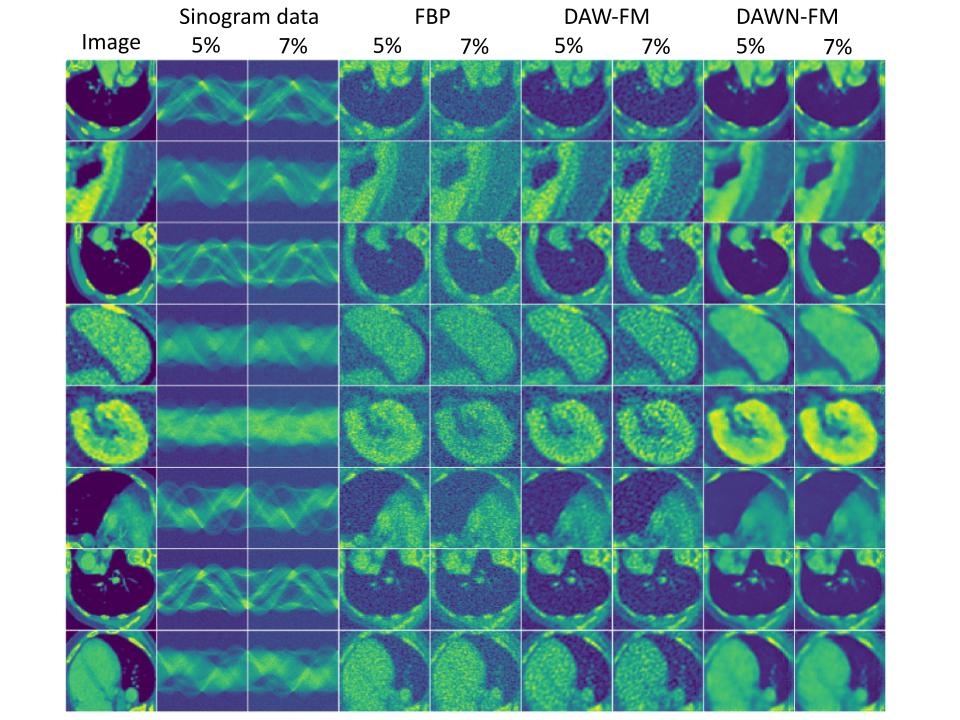}
    \caption{Comparison between different methods the \textbf{tomography} task on the \textbf{OrganAMNIST} dataset for noise levels $p = 5\%$ and 7\% added to the sinogram data.}
    \label{fig:AppendixOrganAMNIST}
\end{figure}

\begin{figure}[h]
    \centering
    \includegraphics[width=0.9\linewidth]{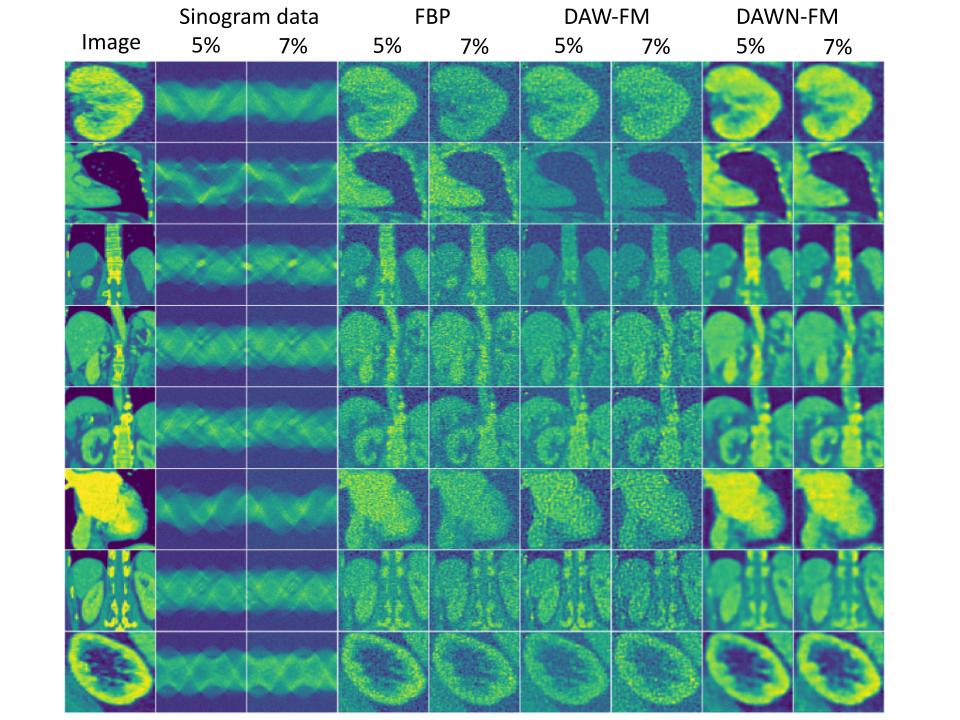}
    \caption{Comparison between different methods the \textbf{tomography} task on the \textbf{OrganCMNIST} dataset for noise levels $p = 5\%$ and 7\% added to the sinogram data.}
    \label{fig:AppendixOrganCMNIST}
\end{figure}

We present the results of our methods \sitd~and \sitdn~for the tomography task and compare them to FBP with ramp filtering in \Cref{tab:tomography_results} at noise level $p = 5\%$ in the sinogram data for OrganAMNIST and OrganCMNIST datasets \cite{medmnistv2}. For this task, \sitdn~outperformed \sitd~on all metrics for both datasets. Some example images for the tomography task for different levels of noise are presented in \Cref{fig:AppendixOrganAMNIST,fig:AppendixOrganCMNIST} in \Cref{app:additional_examples}. In medical image analysis, accurately estimating the size and boundaries of a lobe (such as the one shown in red box in \Cref{fig:medmnist_uncertainty}) is crucial for diagnostics, especially in cases where its size can influence medical decisions. We quantify uncertainty for both \sitd~and \sitdn~by computing mean and standard deviation across 32 samples from the learned posterior. While the mean represents the most probable reconstruction, the standard deviation map highlights regions of higher variability, indicating areas of uncertainty. From \Cref{fig:medmnist_uncertainty}, \sitdn~gives a robust understanding of uncertainty by highlighting boundaries of objects as regions of highest uncertainty, a feature similar to what was observed on the MNIST dataset (see \Cref{fig:uncertainty}).  

\section{Conclusions}
\label{sec:conclusion}
In this paper, we presented \sitdn, a framework for addressing highly ill-posed inverse problems by efficiently incorporating data and noise into the FM process. Our experiments showed that our proposed method consistently outperformed existing methods in tasks like image deblurring, with significant improvements in key performance metrics. The ability to sample from the learned posterior enables the exploration of the solution space and facilitates in uncertainty quantification, which is critical for real-world applications. Future work will focus on refining the model to enhance efficiency and applicability to more ill-posed problems, including integration of advanced noise modeling techniques for extreme noise conditions.

\section*{Acknowledgments}
S. Ahamed would like to acknowledge support from the Canadian Institutes of Health Research (CIHR) Grant PIBH-GR018169. Computational resources were provided by E. Haber.








\bibliographystyle{unsrt}
\bibliography{aims_references}
\medskip
Received February 27, 2025; revised October 19, 2025; early access January 20,
2026.
\medskip

\newpage
\clearpage

\section*{Supplementary Material}
\label{supp_material}

\section{Evaluation Metrics}
\label{app:evaluation_metrics}

In this section, we provide detailed explanation of the evaluation metrics used in our experiments for 2D images, including the Mean Squared Error (MSE), Misfit, Structural Similarity Index Measure (SSIM), and Peak Signal-to-Noise Ratio (PSNR).

\subsection{Mean Squared Error (MSE)}
The Mean Squared Error (MSE) measures the average squared difference between the pixel intensities of the original image and the reconstructed image. Given two images $I_{\text{true}}(x, y)$ (the ground truth image) and $I_{\text{rec}}(x, y)$ (the reconstructed image), MSE is calculated as,

\begin{equation}
    \text{MSE} = \frac{1}{H \cdot W} \sum_{x=1}^{H} \sum_{y=1}^{W} \left( I_{\text{true}}(x, y) - I_{\text{rec}}(x, y) \right)^2,
\end{equation}

where,
\begin{itemize}
    \item $H$ and $W$ are the height and width of the image,
    \item $I_{\text{true}}(x, y)$ is the pixel value at location $(x, y)$ in the ground truth image,
    \item $I_{\text{rec}}(x, y)$ is the corresponding pixel value in the reconstructed image.
\end{itemize}

A lower MSE indicates better reconstruction performance.

\subsection{Misfit}
The Misfit metric measures how well the forward model of the reconstructed image fits the actual observed data. For 2D images, given a forward operator $A$ and observed data $b$, the Misfit is calculated as:

\begin{equation}
    \text{Misfit} = \frac{1}{2} \sum_{x=1}^{H} \sum_{y=1}^{W} \left( \bfA I_{\text{rec}}(x, y) - b(x, y) \right)^2,
\end{equation}

where,
\begin{itemize}
    \item $\bfA$ is the forward operator (e.g., a blurring or projection operator),
    \item $I_{\text{rec}}(x, y)$ is the reconstructed image,
    \item $b(x, y)$ is the observed (blurred or noisy) image.
\end{itemize}

A lower Misfit indicates that the data obtained from the reconstructed image is consistent with the observed data.

\subsection{Structural Similarity Index Measure (SSIM)}
The Structural Similarity Index Measure (SSIM) assesses the perceptual similarity between two images by considering luminance, contrast, and structure. For 2D images, SSIM is defined as:

\begin{equation}
    \text{SSIM}(I_{\text{true}}, I_{\text{rec}}) = \frac{(2 \mu_{\text{true}} \mu_{\text{rec}} + C_1)(2 \sigma_{\text{true,rec}} + C_2)}{(\mu_{\text{true}}^2 + \mu_{\text{rec}}^2 + C_1)(\sigma_{\text{true}}^2 + \sigma_{\text{rec}}^2 + C_2)},
\end{equation}

where,
\begin{itemize}
    \item $\mu_{\text{true}}$ and $\mu_{\text{rec}}$ are the local means of the ground truth and reconstructed images, respectively,
    \item $\sigma_{\text{true}}^2$ and $\sigma_{\text{rec}}^2$ are the local variances of the ground truth and reconstructed images,
    \item $\sigma_{\text{true,rec}}$ is the local covariance between the two images,
    \item $C_1$ and $C_2$ are constants to avoid division by zero.
\end{itemize}

SSIM values range from $-1$ to $1$, where $1$ indicates perfect structural similarity.

\subsection{Peak Signal-to-Noise Ratio (PSNR)}
The Peak Signal-to-Noise Ratio (PSNR) measures the quality of the reconstructed image compared to the ground truth image. Since our images are normalized between 0 and 1, PSNR is defined as:

\begin{equation}
    \text{PSNR} = -10 \cdot \log_{10} (\text{MSE}),
\end{equation}

where, $\text{MSE}$ is the Mean Squared Error between the ground truth and reconstructed images. Higher PSNR values indicate better reconstruction quality.

\section{Additional Experiments with Inverse Crime Considerations}
\label{app:inverse_crime}

To examine whether our results are affected by the inverse crime \cite{wirgin2004inverse}, we conducted additional experiments in which the forward model used for data generation differed from that used during inversion. Specifically, instead of generating synthetic data directly using $\bfA$ as $\bfb = \bfA \bfx_1 + \bfepsilon$, we defined a composite forward operator,
\begin{equation}
    \Tilde{\bfA} = \bfD \bfA \bfU,
\end{equation}
where $\bfU$ is an upsampling operator (by a factor of 2) applied to the clean image $\bfx_1$ and $\bfD$ is a downsampling operator (by a factor of 2) applied after the forward operation. This composite operator introduces a domain mismatch between the forward model used for data generation and the one used for inversion (which remains $\bfA$), thereby eliminating the inverse crime. The results in \Cref{tab:resolveIC} demonstrate that DAWN-FM maintains comparable reconstruction quality even under these more challenging conditions - for example, on the MNIST deblurring task, performance metrics such as MSE, SSIM, and PSNR remain nearly unchanged, and similar robustness is observed for the tomography task on OrganCMNIST. These results empirically confirm that the effectiveness of DAWN-FM seems to be robust to any implicit inverse crime, and the method generalizes well even when the forward model used for data generation is perturbed.

\begin{table}[h]
\centering
\caption{Performance of DAWN-FM for the deblurring (on MNIST) and tomography (on OrganCMNIST) tasks with (using data generation forward problem, $\bfA$) and without (using data generation forward problem, $\bfD \bfA \bfU$) inverse crime. Results show negligible degradation when a composite forward model slightly different from $\bfA$ is used for data generation.}
\label{tab:resolveIC}
\resizebox{0.95\columnwidth}{!}{%
\begin{tabular}{@{}ccccccccc@{}}
\toprule
\multirow{2}{*}{\begin{tabular}[c]{@{}c@{}}\textbf{Task} \\ \textbf{(Dataset)}\end{tabular}} & \multirow{2}{*}{\textbf{Methods}} & \multicolumn{2}{c}{\begin{tabular}[c]{@{}c@{}}\textbf{Forward} \\ \textbf{model}\end{tabular}} & \multirow{2}{*}{\begin{tabular}[c]{@{}c@{}}\textbf{Inverse} \\ \textbf{crime} \\ \textbf{present?}\end{tabular}} & \multirow{2}{*}{\textbf{MSE} ($\downarrow$)} & \multirow{2}{*}{\textbf{MISFIT} ($\downarrow$)} & \multirow{2}{*}{\textbf{SSIM} ($\uparrow$)} & \multirow{2}{*}{\textbf{PSNR} ($\uparrow$)} \\ \cmidrule(lr){3-4}
 &  & \multicolumn{1}{l}{\textbf{Generation}} & \multicolumn{1}{l}{\textbf{Inversion}} &  &  &  &  &  \\ \midrule
\multirow{2}{*}{\begin{tabular}[c]{@{}c@{}}Deblurring \\ (MNIST)\end{tabular}} & \multirow{4}{*}{DAWN-FM} & $\bfA$ & $\bfA$ & Yes & 0.073 $\pm$ 0.041 & 0.032 $\pm$ 0.009 & 0.901 $\pm$ 0.042 & 22.277 $\pm$ 2.612 \\
 &  & $\bfD \bfA \bfU$ & $\bfA$ & No & 0.070 $\pm$ 0.034 & 0.041 $\pm$ 0.095 & 0.870 $\pm$ 0.090 & 22.198 $\pm$ 1.816 \\ \cmidrule(r){1-1} \cmidrule(l){3-9} 
\multirow{2}{*}{\begin{tabular}[c]{@{}c@{}}Tomography \\ (OrganCMNIST)\end{tabular}} &  & $\bfA$ & $\bfA$ & Yes & 0.007 $\pm$ 0.007 & 0.151 $\pm$ 1.287 & 0.675 $\pm$ 0.116 & 21.630 $\pm$ 3.199 \\
 &  & $\bfD \bfA \bfU$ & $\bfA$ & No & 0.009 $\pm$ 0.062  & 0.162 $\pm$ 1.113 &	0.632 $\pm$ 0.175 &	20.489 $\pm$ 3.795  \\ \bottomrule
\end{tabular}
}
\end{table}

\section{Experimental Settings}
\label{app:experiment_setting}

For the image deblurring task, the experiments were conducted on MNIST, STL10 and CIFAR10 datasets. The networks used were Diffusion model, InverseUNetODE as baselines and our proposed methods \sitd~and \sitdn. The key details of the experimental setup for each of these methods are summarized in \Cref{tab:expt_details_deblurring_ourmethods,tab:expt_details_deblurring_diffusion,tab:expt_details_deblurring_inverseunetode}. For the tomography task, the experiments were conducted using \sitd~and \sitdn~methods on OrganAMNIST and OrganCMNIST datasets derived from the MedMNIST data library \cite{medmnistv2}. The key details of the experimental setup are summarized in \Cref{tab:expt_details_tomography_ourmethods}. All our experiments were conducted on an NVIDIA A6000 GPU with 48GB of memory. Upon acceptance, we will release our source code, implemented in PyTorch \cite{paszke2017automatic}.

\begin{table}[]
\centering
\caption{Experimental details for the image deblurring task for our methods \sitd~and \sitdn.}
\label{tab:expt_details_deblurring_ourmethods}
\resizebox{0.95\columnwidth}{!}{%
\begin{tabular}{@{}ll@{}}
\toprule
\textbf{Component} &
  \textbf{Details} \\ \midrule
Datasets &
  \begin{tabular}[c]{@{}l@{}}MNIST, STL10 and CIFAR10. The dimension of images in these \\ datasets were $28\times28$, $64\times64$, $32\times32$, respectively\end{tabular} \\ \midrule
\multirow{2}{*}{Network architectures} &
  \begin{tabular}[c]{@{}l@{}}MNIST and CIFAR10: \\ \textbf{\sitd}: UNet with 3 levels with {[}$c$, 16, 32{]} filters with residual\\ blocks and time and data embedding at each level. \\ Time embed dimension = 256;\\ \textbf{\sitdn}: UNet with 3 level levels with {[}$c$, 16, 32{]} filters with\\ residual blocks and time, data and noise embedding at each level.\\ Time and noise embed dimension = 256.\\ Here $c$ = 1 and 3 for MNIST and CIFAR10, respectively.\end{tabular} \\ \cmidrule(l){2-2} 
 &
  \begin{tabular}[c]{@{}l@{}}STL10: \\ \textbf{\sitd}: UNet with 5 levels with {[}3, 16, 32. 64, 128{]} filters \\ with residual blocks and time and data embedding at each level.\\ Time embed dimension = 256;\\ \textbf{\sitdn}: UNet with 5 levels with {[}3, 16, 32, 64, 128{]} filters\\ with residual blocks and time, data and noise embedding at each \\ level. Time and noise embed dimension = 256\end{tabular} \\ \midrule
\multirow{2}{*}{Number of trainable parameters} &
  \begin{tabular}[c]{@{}l@{}}MNIST and CIFAR10: \\ \textbf{\sitd}: 1,038,398; \\ \textbf{\sitdn}: 1,066,497\end{tabular} \\ \cmidrule(l){2-2} 
 &
  \begin{tabular}[c]{@{}l@{}}STL10: \\ \textbf{\sitd}: 8,009,824;\\ \textbf{\sitdn}: 8,177,955\end{tabular} \\ \midrule
Loss function & MSE loss for the velocity and misfit terms, \Cref{eq:velestMod} \\ \midrule
Optimizer &
  Adam \cite{kingma2014adam}\\ \midrule
\multirow{1}{*}{Learning rate (lr) schedule} &
  \begin{tabular}[c]{@{}l@{}} CosineAnnealingLR with lr$_\text{init}$ = $10^{-4}$, lr$_\text{min}$ = $10^{-6}$ \\and $T_\text{max}$ = max\_epochs \end{tabular} \\ \midrule
\multirow{1}{*}{Stopping criterion} &
  \begin{tabular}[c]{@{}l@{}}\textbf{\sitd}: max\_epochs = 3000; \\ \textbf{\sitdn}: max\_epochs = 3000
  \end{tabular} \\ \midrule
Integrator for ODE &
  Fourth-Order Runga-Kutta with step size $h = 1/100$ \\ \bottomrule
\end{tabular}%
}
\end{table}

\begin{table}[]
\centering
\caption{Experimental details for the image deblurring task for Diffusion model \cite{chung2022diffusion}.}
\label{tab:expt_details_deblurring_diffusion}
\resizebox{0.95\columnwidth}{!}{%
\begin{tabular}{@{}ll@{}}
\toprule
\textbf{Component}                             & \textbf{Details}   \\ \midrule
Datasets &
  \begin{tabular}[c]{@{}l@{}}MNIST, STL10, and CIFAR10. The dimension of images in\\ these  datasets were 28x28, 64x64, 32x32, respectively\end{tabular} \\ \midrule
\multirow{3}{*}{Network architectures} &
  \begin{tabular}[c]{@{}l@{}}MNIST: UNet with 4 levels with {[}1, 16, 32, 64{]} filters and \\ sinusoidal time embedding (1000 time steps) embedded at \\ each level\end{tabular} \\
 &
  \begin{tabular}[c]{@{}l@{}}STL10: UNet with 6 levels with {[}3, 16, 32, 64, 128, 128{]} \\ filters and sinusoidal time embedding (1000 time steps) \\ embedded at each level\end{tabular} \\
 &
  \begin{tabular}[c]{@{}l@{}}CIFAR10: UNet with 4 levels with {[}3, 16, 32, 64{]} filters \\ and sinusoidal time embedding (1000 time steps) \\ embedded at each level\end{tabular} \\ \midrule
\multirow{3}{*}{Number of training parameters} & MNIST: 1,360,712   \\
                                               & STL10: 8,565,086   \\
                                               & CIFAR10: 1,365,598 \\ \midrule
Loss function &
  \begin{tabular}[c]{@{}l@{}} MSE loss between $SNR \cdot x_\text{rec}$ and $SNR \cdot x_1$, \\ where $SNR$ is the signal-to-noise ratio at each time step, $x_\text{rec}$\\ is the reconstructed image and $x_1$ is the original image
  \end{tabular} \\ \midrule
Optimizer                                      & Adam \cite{kingma2014adam}              \\ \midrule
Learning rate (lr)                             & $10^{-4}$ (constant)    \\ \midrule 
Stopping criterion  & max\_epochs = 2000 \\ \bottomrule
\end{tabular}%
}
\end{table}

\begin{table}[]
\centering
\caption{Experimental details for the image deblurring task for InverseUNetODE  \cite{eliasof2024completedeeplearningmethod}.}
\label{tab:expt_details_deblurring_inverseunetode}
\resizebox{0.95\columnwidth}{!}{%
\begin{tabular}{@{}ll@{}}
\toprule
\textbf{Component} &
  \textbf{Details} \\ \midrule
Datasets &
  \begin{tabular}[c]{@{}l@{}}MNIST, STL10, and CIFAR10. The dimension of images in these \\ datasets were 28x28, 64x64, 32x32, respectively\end{tabular} \\ \midrule
Network description &
  \begin{tabular}[c]{@{}l@{}}UNet with each level utilizing a combination of convolutional layer\\ embedding for feature extraction and hyperUNet layer for hierarchical \\ feature refinement. The network incorporates the forward problem by \\ applying the adjoint $\bfA^{\top}$ of forward problem to the residual at each \\ layer to iteratively correct the estimate of the reconstructed image.\end{tabular} \\ \midrule
\multirow{3}{*}{Network architectures} &
  \begin{tabular}[c]{@{}l@{}}MNIST: 3 levels with 3 hidden units per level and 3 nested layers \\ within each hyperUNet layer\end{tabular} \\
 &
  \begin{tabular}[c]{@{}l@{}}STL10:  5 levels with 8 hidden units per level and 3 nested layers \\ within each hyperUNet layer\end{tabular} \\
 &
  \begin{tabular}[c]{@{}l@{}}CIFAR10: 5 levels with 8 hidden units per level and 3 nested layers \\ within each hyperUNet layer\end{tabular} \\ \midrule
\multirow{3}{*}{Number of training parameters} &
  MNIST:  2,089,887 \\
 &
  STL10: 6,503,920 \\
 &
  CIFAR10: 6,503,920 \\ \midrule
Loss function &
  MSE loss between $x_\text{recon}$ (predicted image) and $x_1$ (clean image)  \\ \midrule
Optimizer &
  Adam \cite{kingma2014adam} \\ \midrule
Learning rate (lr) &
  $10^{-4}$ (constant) \\ \midrule
Stopping criterion &
  max\_epochs = 1000 \\ \bottomrule
\end{tabular}%
}
\end{table}

\begin{table}[]
\centering
\caption{Experimental details for the tomography task for our methods \sitd~and \sitdn.}
\label{tab:expt_details_tomography_ourmethods}
\resizebox{0.95\columnwidth}{!}{%
\begin{tabular}{@{}ll@{}}
\toprule
\textbf{Component} &
  \textbf{Details} \\ \midrule
Datasets &
  \begin{tabular}[c]{@{}l@{}}OrganAMNIST and OrganCMNIST from the MedMNIST dataset\\ \cite{medmnistv2}. The dimension of images in both these datasets \\were $64\times64$.\end{tabular} \\ \midrule
\multirow{2}{*}{Network architectures} &
  \begin{tabular}[c]{@{}l@{}}\textbf{\sitd}: UNet with 5 levels with {[}1, 16, 32, 64, 128{]} filters \\ with residual blocks and time and data embedding at each level.\\ Time embed dimension = 256;\\ \textbf{\sitdn}: UNet with 5 levels with {[}1, 16, 32, 64, 128{]} filters\\ with residual blocks and time, data and noise embedding at each \\ level. Time and noise embed dimension = 256\end{tabular} \\ \midrule
\multirow{2}{*}{Number of trainable parameters}  &
  \begin{tabular}[c]{@{}l@{}}\textbf{\sitd}: 8,009,824;\\ \textbf{\sitdn}: 8,177,955\end{tabular} \\ \midrule
Loss function &
  MSE loss for the velocity and misfit terms, \Cref{eq:velestMod} \\ \midrule
Optimizer &
  Adam \cite{kingma2014adam} \\ \midrule
\multirow{1}{*}{Learning rate (lr) schedule} &
  \begin{tabular}[c]{@{}l@{}} CosineAnnealingLR with lr$_\text{init}$ = $10^{-4}$, lr$_\text{min}$ = $10^{-6}$ \\and $T_\text{max}$ = max\_epochs \end{tabular} \\ \midrule
\multirow{1}{*}{Stopping criteria} &
  \begin{tabular}[c]{@{}l@{}}\textbf{\sitd}: max\_epochs = 3000; \\ \textbf{\sitdn}: max\_epochs = 3000
  \end{tabular} \\ \midrule
Integrator for ODE &
  Fourth-Order Runga-Kutta with step size $h = 1/100$ \\ \bottomrule
\end{tabular}%
}
\end{table}

\pagebreak

\section{Visualization}
\label{app:additional_examples}

\end{document}